%% file: main.tex
\documentclass[conference]{IEEEtran}
\usepackage{cite}
\usepackage{amsmath,amssymb,amsfonts}
\usepackage{algorithmic}
\usepackage{graphicx}
\usepackage{textcomp}
\usepackage{xcolor}
\usepackage[hyphens]{url}

\usepackage{fancyhdr}
\usepackage{soul}
\usepackage{algorithm}
\usepackage{xspace}
\usepackage[font=footnotesize,labelfont=bf]{caption}
\usepackage{mathtools}
\usepackage{pifont}
\usepackage{color}

\usepackage{textgreek}
\usepackage{subfig}
\usepackage{textcomp}
\usepackage[]{siunitx}

\usepackage{booktabs}

\def\BibTeX{{\rm B\kern-.05em{\sc i\kern-.025em b}\kern-.08em
    T\kern-.1667em\lower.7ex\hbox{E}\kern-.125emX}}

\newcommand{\old}[1]{}
\newcommand{\fig}[1]{Figure~\ref{#1}}
\newcommand{\sect}[1]{Section~\ref{#1}}
\newcommand{\tab}[1]{Table~\ref{#1}}
\newcommand{\algo}[1]{Algorithm~\ref{#1}}

\newcommand{\nsconfig}[0]{\texttt{NS$_{config}$}\xspace}
\newcommand{\proposed}[0]{SmartSAGE\xspace}

\newcommand{\dram}[0]{DRAM\xspace}
\newcommand{\ssd}[1]{SSD({#1})\xspace}

\newcommand{\sage}[1]{SmartSAGE({#1})\xspace}

\newcommand\blfootnote[1]{%
\begingroup
\renewcommand\thefootnote{}\footnote{#1}%
\addtocounter{footnote}{-1}%
\endgroup
}

\title{\huge SmartSAGE: Training Large-scale Graph Neural Networks\\ using In-Storage Processing Architectures}

\begin{document}

\author{

\IEEEauthorblockN{
Yunjae Lee\hspace{2em}Jinha Chung\hspace{2em}Minsoo Rhu}
\IEEEauthorblockA{
School of Electrical Engineering\\
KAIST\\
\texttt{\{yunjae408, jinha.chung, mrhu\}@kaist.ac.kr}\\
}
}

\maketitle
\thispagestyle{plain}
\pagestyle{plain}


\input{tex/abstract}
\IEEEpeerreviewmaketitle
\blfootnote{
This is the author preprint version of the work. The authoritative version will appear in the Proceedings of the $49^{\text{th}}$ IEEE/ACM International Symposium on Computer Architecture (ISCA-49), 2022.
}
\input{tex/intro}

\input{tex/background}

\input{tex/characterization}

\input{tex/proposed}

\input{tex/methodology}

\input{tex/result}

\input{tex/discussion}
\input{tex/conclusion}

\section*{Acknowledgement}
This research is partly supported by the National Research Foundation of Korea (NRF) grant funded by the Korea
government(MSIT) (NRF-2021R1A2C2091753), the Super Computer Development Leading Program of the NRF funded by the Korea
government MSIT under grant NRF-2020M3H6A1085498, and by Samsung Electronics Co., Ltd (IO201210-07974-01).
We also appreciate the support from Samsung Advanced Institute of Technology (SAIT) and the EDA tools supported by
the IC Design Education Center (IDEC), Korea. Minsoo Rhu is the corresponding author.


\bibliographystyle{IEEEtranS}
\bibliography{refs}

\end{document}

%% file: tex/abstract.tex
\begin{abstract}

Graph neural networks (GNNs) can extract features by learning both the
representation of each objects (i.e., graph nodes) and the relationship
across different objects (i.e., the edges that connect nodes), achieving
state-of-the-art performance in various graph-based tasks.  Despite its
strengths, utilizing these algorithms in a production environment faces several
challenges as the number of graph nodes and edges amount to several
billions to hundreds of billions scale, requiring substantial storage space for
training. Unfortunately, state-of-the-art ML frameworks employ an in-memory
processing model which significantly hampers the productivity of ML
practitioners as it mandates the overall working set to fit within DRAM
capacity.  In this work, we first conduct a detailed
characterization on a state-of-the-art, large-scale GNN training algorithm,
GraphSAGE. Based on the characterization, we then explore the
feasibility of utilizing capacity-optimized NVMe SSDs for
storing memory-hungry GNN data, which enables large-scale
GNN training beyond the limits of main memory size. Given
the large performance gap between DRAM and SSD, however,
blindly utilizing SSDs as a direct substitute for DRAM leads
to significant performance loss.  We therefore develop 
\proposed, our software/hardware co-design based on an
in-storage processing (ISP) architecture.  Our work
demonstrates that an ISP based large-scale GNN training system
can achieve both high capacity storage and high performance,
opening up opportunities for ML practitioners to train large
GNN datasets without being hampered by the physical
limitations of main memory size.
\end{abstract}

%% file: tex/intro.tex
\section{Introduction}
\label{sect:intro}

Deep neural network (DNN) based machine learning (ML) algorithms are
providing super-human performance in areas of image
classification,  natural language processing, speech recognition, and others.
However, such DNN based ML design paradigms primarily targeted Euclidean data
(e.g., image, text, and audio), having limited adoption
in domains where \emph{non}-Euclidean data structures
such as \emph{graphs} are utilized.  Recently, \emph{graph neural networks}
(GNNs) have emerged as a powerful tool in application domains that target
arbitrarily structured graph inputs, where feature vectors are associated with
graph nodes and edges. GNNs have found significant success in the areas of
e-commerce and advertisement where the graph nodes and edges represent objects
and their relationships, providing unparalleled performance in traditional graph
analytics workloads. For instance, Pinterest's PinSAGE~\cite{pinsage} or Alibaba's
AliGraph~\cite{alibaba} leverages GNNs to analyze and extract high quality features
from graphs with billions of user/item feature embeddings, achieving
state-of-the-art performance.

	With the proliferation of GNNs, we are witnessing a
	large number of ML frameworks tailored for graph learning being developed,
	examples of which include Deep Graph Library (DGL)~\cite{dgl} and PyTorch
	Geometric (PyG)~\cite{pyg}.  Unlike conventional DNNs (e.g., convolutional,
			recurrent, or fully-connected layers) which exhibit a highly regular and
	dense dataflow, GNN training inherently contains a \emph{hybrid} mix of both
	sparse and dense dataflows (\fig{fig:gnn_pipe}). More concretely, the
	frontend stages of GNN training which conduct ``input data preparation''
	(e.g., graph neighbor sampling, feature table lookup) follow the typical graph
	analytics' sparse and irregular memory access characteristics. In contrast,
	the backend stages which conduct ``graph learning'' (e.g., graph
			convolutions) employ well-structured, dense DNN algorithms using
	multi-layer perceptron (MLP) layers.  Therefore, DGL and PyG provides
	dedicated user-level APIs tailored to the sparse dataflow of input data
	preparation stages (i.e., fine-grained feature gather-scatter for graph
			neighbor sampling and feature aggregation) which eases the programming of
	irregular and sparse frontend stages of GNN training.

	While DGL and PyG help improve the programmability of GNNs,
	both frameworks employ the \emph{in-memory} processing model~\cite{pyg,dgl} (i.e., 
the target graph
			nodes/edges and its feature embedding tables must be stored inside main
			memory) which limits ML practitioners from scaling ``up'' the graph dataset, 
	significantly hampering user productivity.
	General trend in the GNN
	research space has been to increase the number of graph nodes as well as the number of edges
	(i.e., larger and more complex graphs) while employing more sophisticated DNNs
	for feature extraction (i.e., convolutions~\cite{kipf:semi_supervised_gcn:iclr2017} to attentions~\cite{velickovic:gan:iclr2018}).
	An important reason why such scaled-up GNNs became prevalent is because it
	helps improve GNN's algorithmic performance, similar to how ML 
	algorithms targeting Euclidean data show higher performance with larger and
	deeper DNNs.  
	Unfortunately, current ML frameworks' in-memory processing
	model forces ML practitioners
	to tune the GNN training algorithm to fit within 
	 the several tens to hundreds of
	GBs of CPU memory, preventing developers from scaling up the
	graph network structure as well as its feature vector size. 
	To this end, this paper explores the feasibility of utilizing NAND flash-based
	non-volatile memory (NVM) solutions to address the memory ``capacity''
	bottlenecks of large-scale GNN training.  Our proposal encompasses
	innovations at both the software and hardware stack, as detailed below.

{\bf Software.} Given the wide performance gap between DRAM and NVMe SSDs, the
central research challenge lies in how system architects should go about
harmoniously architecting the memory-storage hierarchy that is most appropriate
for GNN training's algorithmic properties.  Consequently, we start by
conducting a chacterization on a state-of-the-art large-scale GNN training
algorithm, GraphSAGE~\cite{graphsage}.  Our characterization reveals that GNN
training's frontend input data preparation stage (\fig{fig:gnn_pipe}) is the
most memory capacity intensive and becomes a prime candidate to be offloaded to
SSDs. We therefore establish a baseline SSD-centric training system which
stores the memory-hungry data structures (i.e., graph nodes/edges) inside SSDs.
These data structures are mapped to user-space memory address via memory-mapped
(mmap) file I/O, which allows the most recently accessed pages to be buffered inside
the OS managed page cache (i.e., stored in DRAM), potentially narrowing the
large performance gap between DRAM and SSDs.  

Unfortunately, our baseline
SSD-centric training system is shown to benefit little from the
locality-optimized OS page cache, incurring high performance loss when
offloaded to SSDs with an average $9.8\times$ slowdown vs. an oracular,
					in-memory processing based system (i.e., all graph datasets can be
							stored in DRAM).  Careful examination of the bottlenecks caused
					by the SSD-offloaded data preparation reveals the following {\bf key
						observation}: because data preparation exhibits fine-grained
						irregular parallelism, it becomes challenging for the page cache to
						reap locality benefits, only to add several tens of microseconds of
						latency in traversing through the system software stack to maintain
						the page cache, rendering the frontend data preparation stage to be
						highly latency limited.  As such, our proposal designs the software
						architecture to be optimized for latency, rather than locality.
						More concretely, we restructure ML framework's software runtime
						system as well as the host driver stack to \emph{bypass} the OS
						system software layers and directly access the SSD without page
						caching.  Such design point obviates the latency overheads in
						maintaining the page cache to buffer recently accessed data,
						significantly reducing the time taken to fetch graph datasets from
						the SSD. 	We demonstrate that optimizing the
						DRAM$\leftrightarrow$SSD data movements for latency leads to
						significant speedup on end-to-end GNN training time with an average
						$2.5\times$ vs. the baseline mmap-based SSD, closing the
						performance gap between DRAM vs. SSDs to ``only'' an average
						$3.8\times$.

{\bf Hardware.} To further close DRAM-vs-SSD's remaining performance gap, we
also innovate at the hardware architecture level driven by both recent
technological trends and GNN algorithm awareness.  Recent trends point to the
emergence of SSD storage devices that natively support \emph{in-storage
	processing} (ISP) capabilities (e.g., 		Samsung-Xilinx's
			SmartSSD~\cite{smartSSD}, NGD system's Newport~\cite{newport,newport_product_webpage}, Eideticom's NoLoad
			CSP~\cite{NoLoadCSP}) that offer fast communication between the flash devices
	and the ISP units. 
	The key objective of our proposal is to synergistically combine ISP capabilities of these
	emerging computational storage devices (CSDs) with our latency-optimized
	software runtime system and host driver stack to achieve ``DRAM-level'' effective 
	throughput, as provided with the in-memory processing baseline ML frameworks.  
	
	To
	this end, we develop \proposed, an ISP based GNN training system that
	intelligently offloads the data intensive stages to the ISP unit,
	closely coupled inside the SSD. The key research problem naturally lies in
	identifying which part of the GNN training algorithm is most appropriate to
	be handled by our ISP architecture.  Our characterization reveals that the
	nature of GNN training's data preparation is effectively a \emph{reduction}
	operation as it seeks to extract out multiple ``subgraphs'' from a much
	larger input graph. 
	Rather than having large, coarse-grained chunks of the input graph be transferred 
	from SSD to DRAM for	subgraph generation by the host CPU,
	\proposed offloads 
  the data intensive steps of data preparation (more specifically, the \emph{neighbor sampling}
	step) to the ISP units. This allows
	\proposed to
	only transfer the subgraphs from SSD to DRAM, preprocessed by the ISP unit, 
	significantly reducing the data movements between SSD$\rightarrow$DRAM by
	an average $20\times$.  Putting everything together, \proposed
	holistically combines our latency-optimized software system with an ISP
	accelerated hardware architecture, achieving substantial speedup on end-to-end GNN training time with an average $3.5\times$ (max $5.0\times$) vs. the baseline SSD-centric system.	
 To summarize our key contributions: 

	\begin{figure}[t!] \centering
\includegraphics[width=0.47\textwidth]{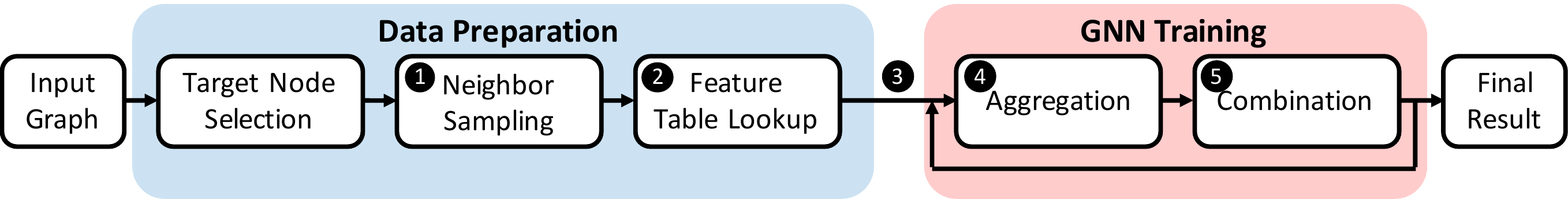}
\caption{High-level overview of a large-scale GNN training pipeline.}
\vspace{-.5em}
\label{fig:gnn_pipe}
\end{figure}

	\begin{itemize}
	\item We conduct a detailed characterization
	on the data-intensive frontend data preparation stage of large-scale GNN training, root-causing
	several limitations of conventional in-memory processing GNN training systems.
	\item  
	Driven by our characterization, we motivate and explore the viability of
	exploiting NVMe SSDs as a direct substitute for capacity limited DRAM based
	training systems.

	\item To bridge the wide performance gap between DRAM vs. SSD, we co-design
	the software/hardware of large-scale GNN training, presenting
	our latency-optimized software coupled with an in-store processing
	architecture, achieving superior performance than baseline SSD-centric systems.

	\end{itemize}

%% file: tex/background.tex
\section{Background}
\label{sect:background}

\subsection{Graph Neural Networks}
\label{sect:gnn}

GNNs are a variant of DNNs that operate over graph data structures. A unique
property of GNNs is that they try to extract features by learning both the
representation of each objects (i.e., graph nodes) as well as the relationship across
different objects (i.e., the edges that connect nodes). Consider the task of
recommending a video/movie clip in an online video streaming service (e.g., YouTube, Netflix). A purely content based approach (e.g., recurrent neural
			networks~\cite{reference_3_from_pinsage_paper}) would represent each
		object under consideration (i.e., the movie) based on the features derived
		from the target object (e.g., the genre of the movie). However, there can
		be valuable information existent across two (and potentially many) distinct
		objects, as there could be common properties that these objects share. By
		modeling such relationship as graph edges and each individual objects as
		graph nodes, a GNN can \emph{learn} to extract meaningful feature
		representations not only from the object itself but also from the edges
		that are associated with it.
	
		Consequently, GNN-based ML applications are achieving state-of-the-art
		performance on a wide range of graph-based algorithms, which
		include node/edge prediction~\cite{kipf:semi_supervised_gcn:iclr2017,
			duvenaud:gcn_molecular:nips2015,
			fout:protein_interface_prediction:nips2017}, graph
			clustering~\cite{ying:graph_clustering:nips2018}, recommendation
			models~\cite{dai2018learning}, and others.  Graph convolutional neural networks for
			instance aggregate target nodes' features and context information from
			their \emph{k}-hop neighborhood nodes, similar to how conventional 
			convolutional layers of a DNN application extracts features from a given pixel's
			neighborhood pixels.  By stacking multiple of such convolution operations
			in sequence (e.g., two convolution layers consider up to two-hop
					(\emph{k}$=$$2$) neighborhoods), the coverage of feature learning
			could exponentially propagate far reaches of the graph.

\begin{figure}[t!] \centering
\includegraphics[width=0.47\textwidth]{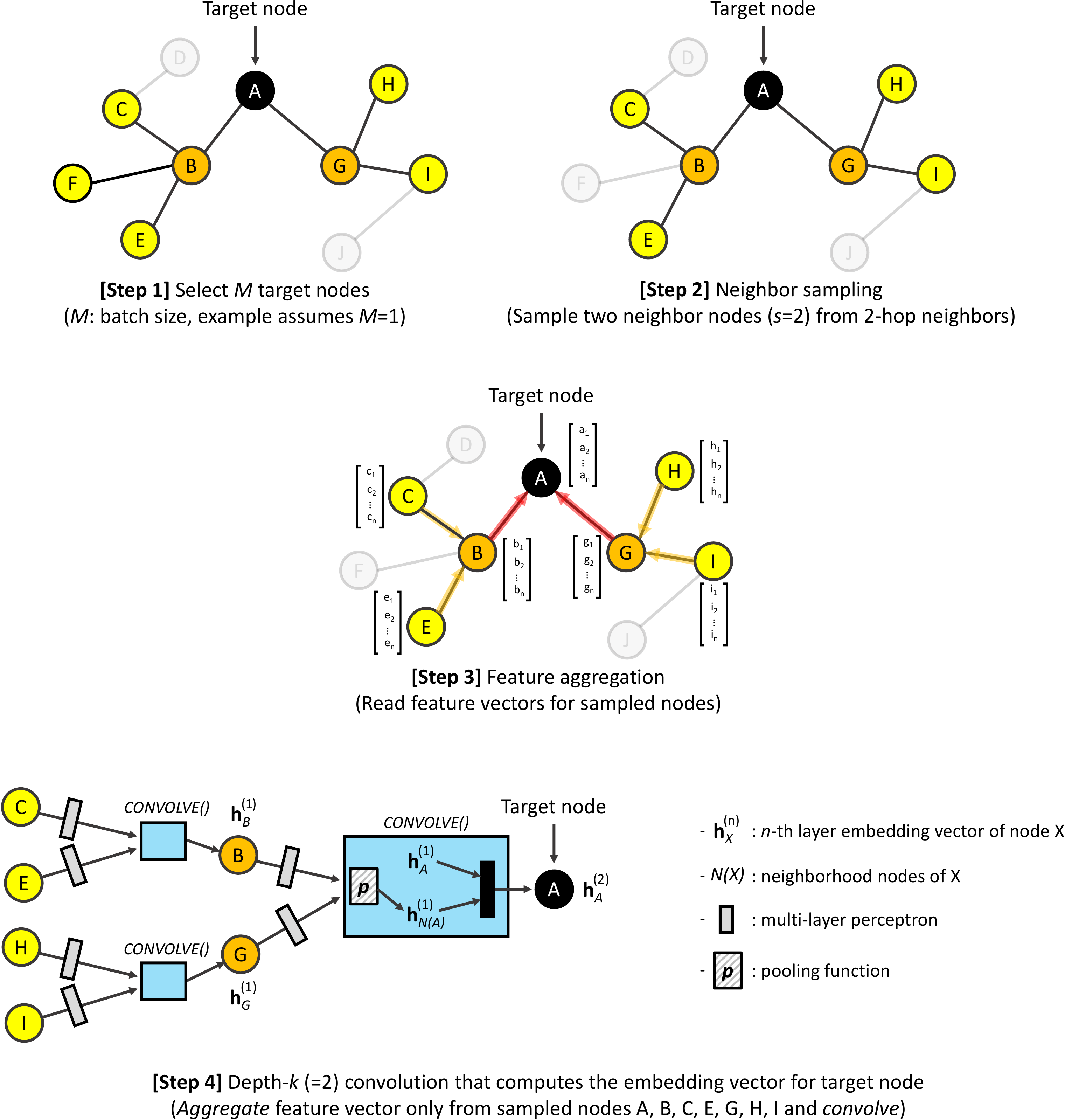}
\vspace{0.5em}
\caption{
Illustration of the GraphSAGE sample and aggregate operator generating a single target node's embedding vector through a depth-$2$ convolution (considering upto two-hop (\emph{k}$=$$2$) neighboring nodes).
Training GNNs with a mini-batch of $M$ involves selecting $M$ target nodes (i.e., $M$=$1$
		in this figure), each going through the sample and aggregation process. Example assumes
two neighboring nodes are sampled per each target node (i.e., \emph{s}$=$$2$). For instance,
	among the three neighbor nodes of node $B$ (nodes $C$/$F$/$E$ in step $1$), we only include nodes $C$ and $E$ (step $2$)
as part of the subgraph. In this work, we focus on addressing the bottlenecks incurred in step $1$ and $2$.
}
\vspace{-.5em}
\label{fig:gnn_101}
\end{figure}

\subsection{``Sample and Aggregate'' for Large-scale GNN Training}
\label{sect:graphsage}

While GNNs have established a new standard in a variety of applications 
within the academic community,
			 they have only recently started being deployed in practical, real-world
			 problems~\cite{pinsage,alibaba}. A key challenge that production
			 environments face is that the number of graph nodes and edges amount to
			 several billions to hundreds of billions scale, requiring substantial
			 compute and memory for training and deployment.  Training early GNN
			 models~\cite{kipf:semi_supervised_gcn:iclr2017} required operating on the
			 \emph{full} graph Laplacian during training, so the \emph{entire} graph data and
			 intermediate states of all graph nodes must be stored in memory. Such
			 high compute and memory requirements can only be met when the target
			 graph data structure is at a small scale.

	 \begin{algorithm}[t!] 
\footnotesize
	 \caption{Neighbor sampling for ``subgraph'' generation}
	 \label{algo:sampling}
	 \begin{algorithmic}[1] 
	 \STATE Set of target nodes $M$;
	 neighborhood $N$; sampling size $s$; Sampled set of nodes $S$ \STATE \STATE
	 /* \textbf{Sampling} neighborhood nodes of target nodes */ \STATE $S
	 \leftarrow \emptyset$ \FOR {$u \in M$} \FOR {$i \leftarrow$ $0$ to $s$}
	 \STATE /* Randomly select its neighborhood nodes */ \STATE $v \leftarrow
	 \texttt{RandomSelect}(N(u))$ \STATE $S \leftarrow S \cup {v}$ \ENDFOR
	 \ENDFOR 
	 \end{algorithmic} 
	 \end{algorithm}

	 To address the practical needs of scaling GNN training and deployment to
	 massive, ``web-scale'' graph data, the seminal work on GraphSAGE (short for
			 graph \emph{sampling} and \emph{aggregation}~\cite{graphsage,pinsage})
	 proposed a highly scalable GNN training framework enabling large-scale graph
	 learning (\fig{fig:gnn_pipe}).  Rather than targeting the entire graph nodes
	 and all the accompanying neighbor nodes for training, GraphSAGE first
	 chooses a fixed number of $M$ \emph{target nodes} ($M$ being equivalent to
			 the training mini-batch size, typically in the range of several
			 thousands), which is much smaller than the number of entire nodes,
	 e.g., $M$=$1$ in step $1$ of \fig{fig:gnn_101}.  For each target node,
	 GraphSAGE then ``samples'' \emph{s} neighborhood nodes around  each target
	 node (\algo{algo:sampling}, step $2$ of \fig{fig:gnn_101}), and only those
	 sampled nodes near the target nodes will later be targeted for 
	 GNN training (i.e., the aggregation stage followed by combination in
			 \fig{fig:gnn_pipe}).  In effect, GraphSAGE dynamically constructs a
	 \emph{subgraph} (generated over a mini-batch of  $M$ target nodes, each
			 target node containing \emph{s} sampled nodes among its neighbors) and
	 iteratively conducts convolutions only around the subgraph  (step $3$ and
			 $4$ in \fig{fig:gnn_101}).  This allows the total number of target nodes
	 for mini-batch training as well as the number of neighboring nodes for a
	 given target node all be hyperparameters of the training algorithm, which
	 helps drastically reduce the \emph{active} compute and memory requirements
	 of GNN training.  As a result, GNNs are gradually seeing real-world
	 adoption in consumer facing products, a well-known example being the usage
	 at Pinterest (through a GNN called PinSAGE~\cite{pinsage}) to generate
	 embedding feature vectors for images/etc. Given their wide adoption 
	 and general applicability  in enabling large-scale GNN training,

\subsection{System Architecture for GNN Training}
\label{sect:sysarch}

A GNN training algorithm's graph dataset consists of several key data
structures including the graph neighbor edge list array that encapsulates the
	structure of the graph using its adjacency matrix, and the feature table
		which abstracts each node's unique property as a feature vector.  In
		real-world graphs, the number of graph edges outweighs the number of graph
		nodes, so the overall memory consumption is generally dominated by the neighbor edge list array (discussed further 
		in \sect{sect:proposed_hw}/\fig{fig:sampling}) rather than the
		feature table, the aggregate size of which can amount to several hundreds
		to thousands of GBs for large-scale graphs.  Such constraint poses several
		key challenges in designing the overall system architecture for large-scale
		GNN training.

Consequently, state-of-the-art GNN training systems typically employ a hybrid CPU-GPU design 
where the frontend data preparation is undertaken by the CPU while the backend
GNN training is handled by the GPU (\fig{fig:gnn_training_system}(a)).  Because
GPUs employ bandwidth-optimized but capacity-limited HBM, they are unable to
locally store the memory capacity limited neighbor edge list array.  Therefore,
				capacity-optimized CPU DIMMs are utilized for storing the
				memory-hungry graph data and the CPU goes through the neighbor sampling
				phase using the neighbor edge list array (step \ding{182} in
						\fig{fig:gnn_pipe} and \fig{fig:gnn_training_system}(a)).  Once
				the sampled subgraph is generated, the feature table is looked up to
				aggregate the corresponding feature vectors of each sampled node (step
						\ding{183}).  The aggregated feature vectors are then copied over
				to the GPU over PCIe (step \ding{184}) for GNN training (step
						\ding{185} and \ding{186}).

%% file: tex/characterization.tex
\begin{figure}[t!] 
\centering
\subfloat[]{
\includegraphics[width=0.23\textwidth]{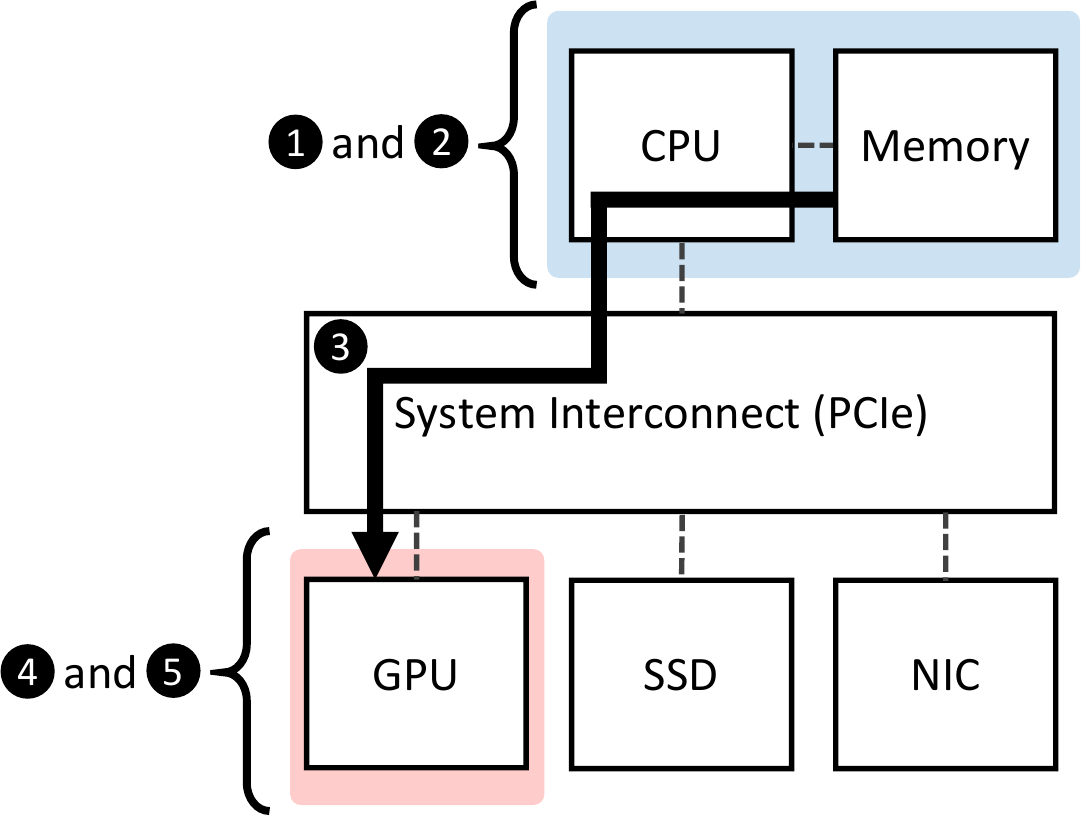}
	\label{fig:}
}
\subfloat[]{
	\includegraphics[width=0.23\textwidth]{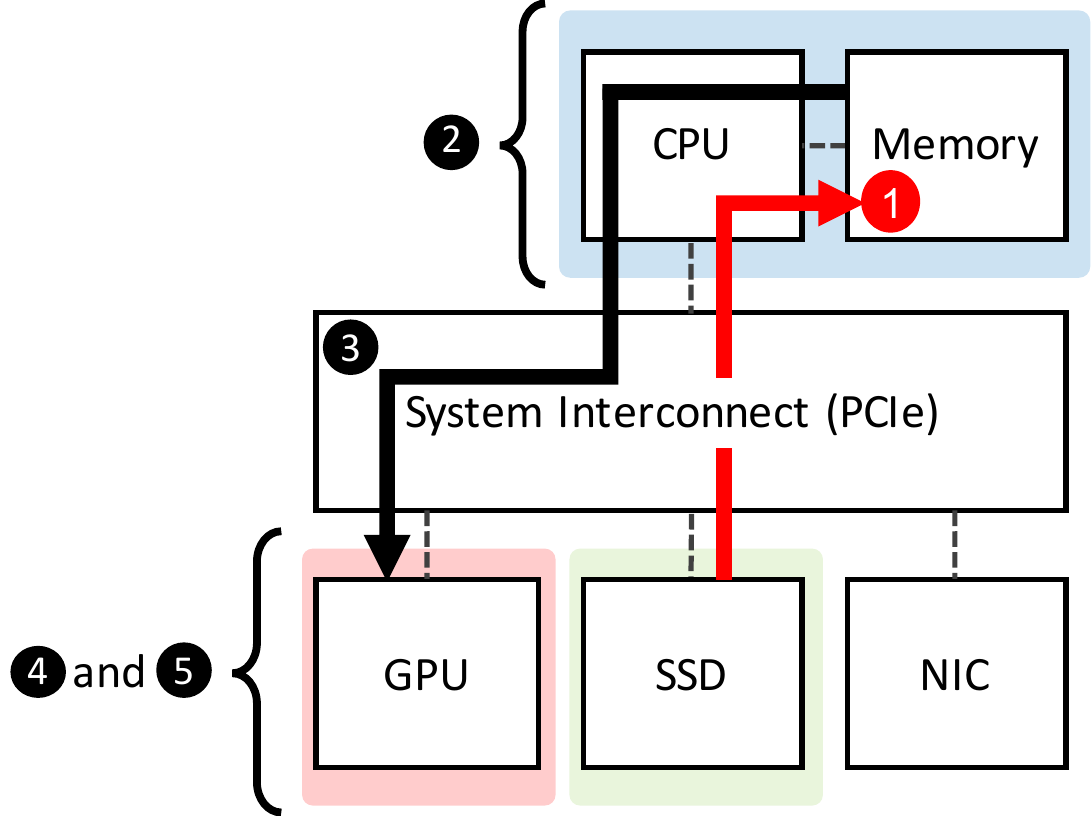}
	\label{fig:}
}
\caption{ 
	(a) The hybrid CPU-GPU based GNN training system assuming an in-memory
		processing model.  (b) The baseline SSD-centric training system utilizes NVMe SSDs 
		to store the
		memory-hungry neighbor edge list array which
		forces the neighbor sampling stage (step \ding{182}) to be conducted over
		the SSD, rather than DRAM, causing performance overheads. Note that step \ding{182}
	to \ding{186} in this figure each matches the corresponding steps \ding{182}$-$\ding{186}
	in \fig{fig:gnn_pipe}.
}   
\label{fig:gnn_training_system}
\vspace{-.5em}
\end{figure}

%
\section{Motivation and Characterization}
\label{sect:characterization}

\subsection{Motivation}
A critical limitation with current graph learning frameworks (DGL~\cite{dgl} and
		PyG~\cite{pyg}) is that their \emph{in-memory} processing model (i.e., the
			key data structures of large-scale GNN training must \emph{all} be stored
			in DRAM) prevents developers from scaling up the graph structure and its
		feature vector size.  One promising alternative is to employ NVMe SSDs for
		storing the memory-hungry data structures of GNN training
		(\fig{fig:gnn_training_system}(b)), utilizing main memory	as a fast cache
		for high locality data.  A potential challenge with SSDs, however, is that
		they operate as block devices where data is transferred in coarse $4$ KB
		chunks at a much lower throughput than DRAM. 
		In the
		remainder of this section, we conduct a detailed characterization on
		GraphSAGE based large-scale GNN training and identify key challenges of utilizing SSDs to
		address the memory capacity limitations of current in-memory processing ML frameworks.

\subsection{Data Preparation in In-Memory Training}
\label{sect:characterization_in_memory}

\begin{figure}[t!] 
\centering
\subfloat[]{
	\includegraphics[width=0.47\textwidth]{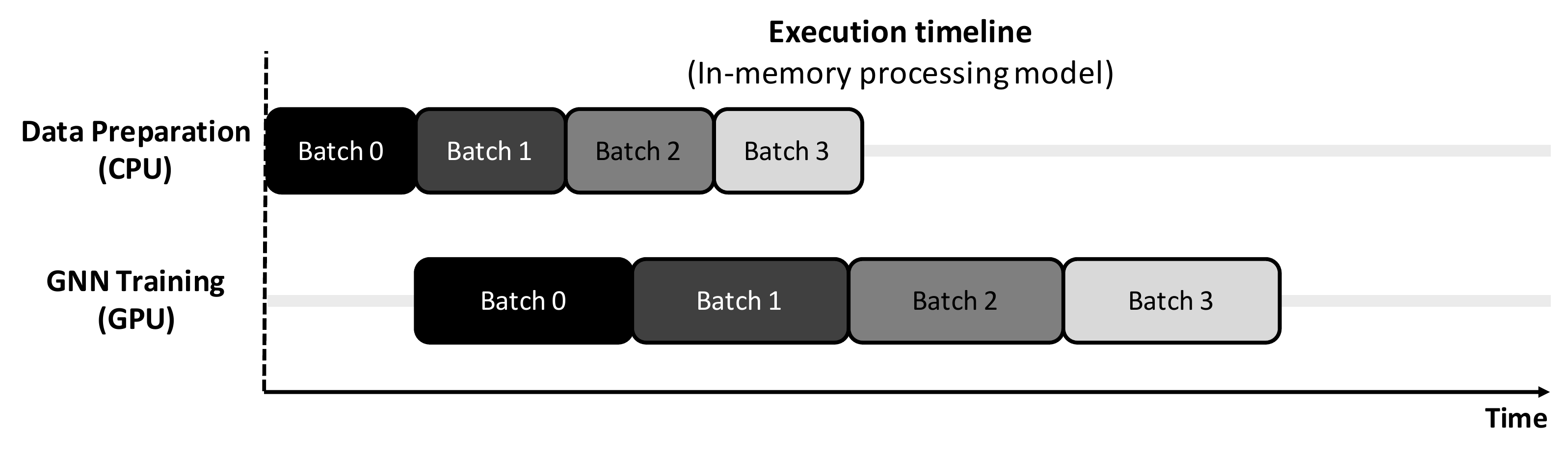}
	\label{fig:execution_timeline_model_in_memory}
}
\vspace{0.1em}
\subfloat[]{
	\includegraphics[width=0.47\textwidth]{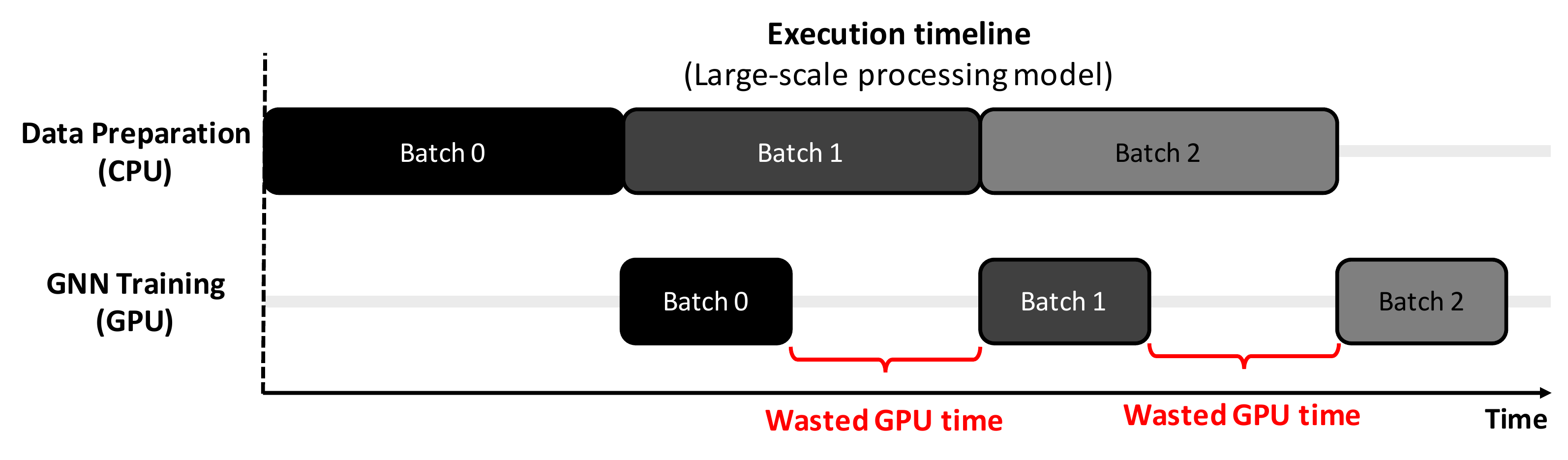}
	\label{fig:execution_timeline_model_terabyte_scale}

} 
\vspace{0.1em} 
\caption{Execution timeline of hybrid CPU-GPU 
	under (a) the baseline system assuming in-memory processing and (b) when the SSD
		is used to store the large-scale graph data. A single subgraph is generated by
		a CPU-side producer worker process, multiples of which are stored into the GPU
		work queue to be 
		consumed by the GPU worker process sequentially for GNN training.
}
\label{fig:execution_timeline_model}
\end{figure}

A typical GNN training pipeline employs the producer-consumer model as illustrated
in \fig{fig:execution_timeline_model}.  The data preparation stage is
implemented using multiple CPU-side producer workers operating in parallel,
						each of which conducts neighbor sampling to generate
						the mini-batch inputs, i.e., the subgraphs. These subgraphs are
						then stored into a work queue which the GPU-side consumer process
						utilizes to initiate GNN training.

The implication of storing the memory capacity limited graph datasets in an
SSD is that the neighbor sampling stage of data preparation (step
		\ding{182} in \fig{fig:gnn_pipe}, \fig{fig:gnn_training_system}(b)) must
now be conducted over the slow and low bandwidth I/O block interface.
Understanding the algorithmic property of sampling and its memory access
behavior is therefore crucial in gauging the feasibility of SSDs for
large-scale GNN training. 

\begin{figure}[t!] \centering
\includegraphics[width=0.475\textwidth]{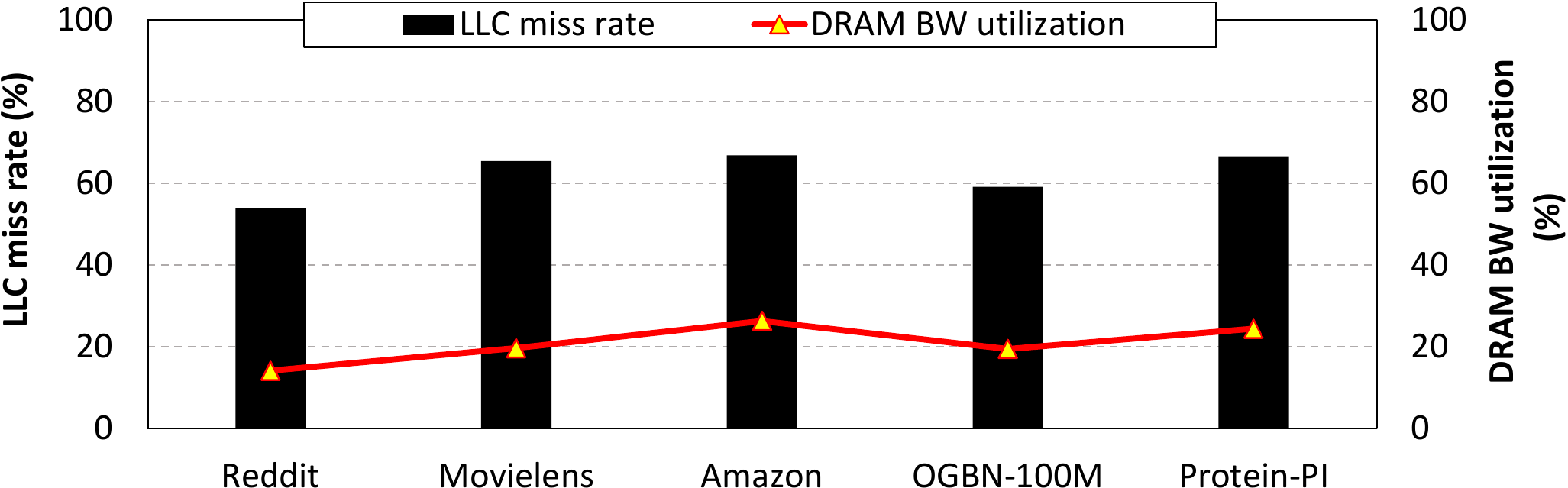}
\caption{The LLC miss rate (left) and DRAM bandwidth utilization (right) during the neighbor sampling stage with baseline in-memory processing training using PyG. We utilize
Linux \texttt{perf} (caching) and Intel RDT utility (bandwidth) for our evaluation.}
\vspace{-0.5em}
\label{fig:bw_utilization}
\end{figure}

To this end, in \fig{fig:bw_utilization}, we start by first characterizing
the on-chip caching and off-chip DRAM bandwidth utilization of the neighbor sampling
algorithm under the baseline in-memory processing setting (i.e., all input graph
		dataset is assumed to be entirely stored in DRAM).  Results show that
neighbor sampling exhibits low caching efficiency with an average $62\%$
last-level cache (LLC) miss rate. As discussed in \algo{algo:sampling},
	neighbor sampling involves randomly selecting a fixed number of nearby nodes
	for each of the target nodes as it traverses through the \emph{k}-hop
	neighborhoods.  Because the target nodes that constitute a training
	mini-batch are typically scattered across the input graph, the execution
	of neighbor sampling is dominated by (random) memory lookup operations
	with little compute intensity, exhibiting a highly sparse and irregular
	dataflow.  Interestingly, the off-chip memory bandwidth utilization is
	generally low despite such high memory intensity, consuming only an average
	$21\%$ of $125$ GB/sec maximum memory throughput. This is because each
	sampling operation only amounts to a fine-grained $8$ byte read
	transaction, leading to severe underutilization of DRAM read throughput.
	Such characterization result implies that the neighbor sampling algorithm is
	severely memory latency limited, rather than throughput limited,
	providing guidelines on optimizing our SSD based training system.  We now
	explore the implication of conducting neighbor sampling over an SSD.

\subsection{Data Preparation in SSD-centric Training}
\label{sect:characterization_ssd}

We establish our baseline SSD-centric training system 
to store the memory
capacity limited graph dataset (esp. the neighbor edge list array) inside the SSD for
the neighbor sampling operation (\fig{fig:gnn_training_system}(b)). These data
structures are accessed using memory-mapped file I/O (mmap) which maps the contents
of the graph dataset file within the user-space memory address. Because the
most recently accessed pages are buffered inside main memory's OS page cache,
		 it is possible for our baseline SSD-centric system to significantly reduce the
		 data fetch latency for high locality accesses, narrowing the performance
		 gap between DRAM vs. SSD.

		 \begin{figure}[t!] \centering
\includegraphics[width=0.475\textwidth]{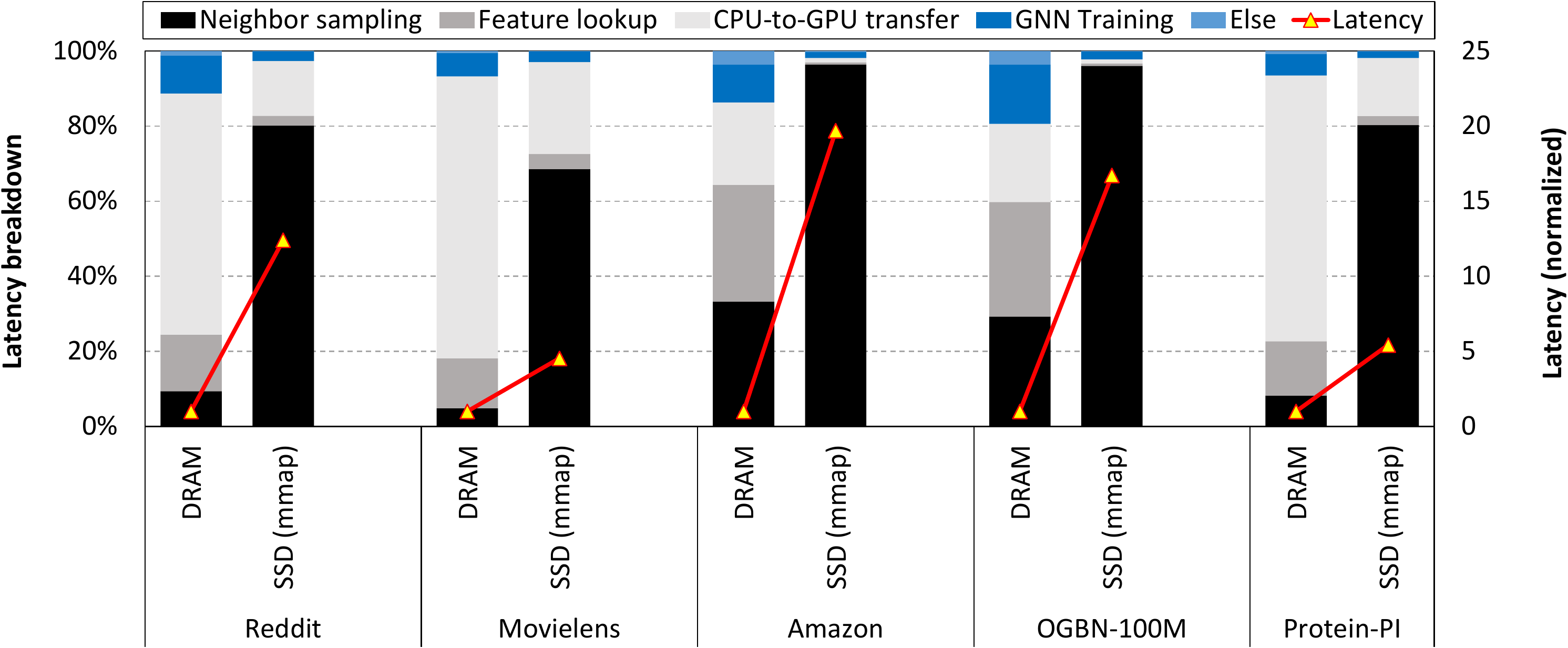}
\caption{
	(Left-axis) Breakdown of GNN training time into key steps of data preparation
		(black-gray) and GNN training (blue). (Right-axis) End-to-end training time
		normalized to the baseline in-memory processing system (\dram). Experiments are conducted
using PyG.}
\vspace{-0.5em}
\label{fig:latency_breakdown}
\end{figure}

\begin{figure}[t!] \centering
\includegraphics[width=0.475\textwidth]{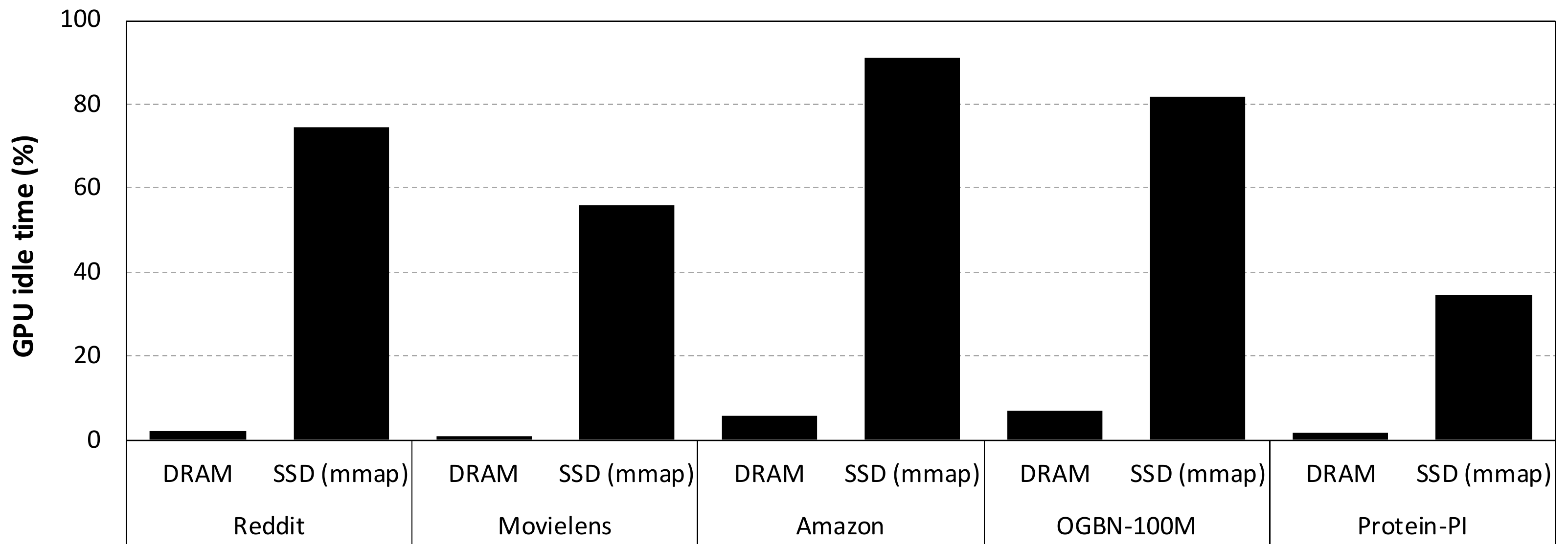}
\caption{Percentage of training time where the GPU is idle due to the lack of input mini-batches to process.}
\vspace{-0.5em}
\label{fig:interval_time}
\end{figure}

\fig{fig:latency_breakdown} provides a breakdown of end-to-end GNN training time
(left-axis) as well as normalized latency (right-axis) when comparing the
baseline in-memory processing vs. mmap-based SSD-centric training system.  As
depicted, the baseline SSD-centric system incurs an average $9.8\times$
(maximum $19.6\times$) slowdown vs. in-memory processing.  As discussed
in \sect{sect:characterization_in_memory}, neighbor sampling exhibits
fine-grained irregular parallelism with low locality. Therefore, the
mmap-based SSD design point experiences significant misses in the OS page cache
during neighbor sampling, so the merits of utilizing the page cache to reap
locality benefits are outweighed by the high latency overheads of maintaining
the OS managed page cache itself (e.g., handling the page-faults incurred
		during the memory-mapped neighbor edge list array accesses, bringing in the
		faulted pages into the page cache, frequent user-kernel space context
		switches, etc).  Consequently, the baseline SSD-centric system suffers from
a  throughput mismatch between the producer-consumer, leaving the GPU idle
whenever the work queue runs out of subgraphs to use as inputs for GNN
training.  In \fig{fig:interval_time}, we quantify the magnitude of such
producer-consumer throughput mismatch by showing the fraction of training time
where the GPU is left idle, waiting for the subgraphs to be generated by the
CPU-side producer workers.  The baseline in-memory processing model shows
high GPU utilization because the data preparation stage is capable of
generating input subgraphs at high throughput
(\fig{fig:execution_timeline_model}(a)).  With the baseline mmap-based
SSD-centric system, however, the producer workers fall short in sufficiently
providing large enough number of subgraphs for the GPU to consume, exhibiting
large periods of GPU idle time, causing a significant slowdown
(\fig{fig:execution_timeline_model}(b)). 

Driven by our characterization, this paper explores an ISP based
SSD-centric architecture for large-scale GNN
training.  As we detail in the next section, our
proposition  holistically addresses
the dual challenges of memory capacity limited
large-scale graph learning and the wide
performance gap between DRAM vs. SSD.  

%% file: tex/proposed.tex
\section{SmartSAGE Architecture}
\label{sect:proposed}

\subsection{Architecture Overview}
\label{sect:proposed_overview}

Our proposed \proposed architecture employs an ISP accelerator tailored for subgraph generation,
	co-designed with our latency-optimized software runtime system and host driver. 
This section takes
a bottom-up approach in presenting \proposed, describing our hardware level
innovations first (\sect{sect:proposed_hw}) followed by a discussion of the software architecture that interfaces
our proposed ISP architecture to the ML frameworks (\sect{sect:proposed_sw}).

\begin{figure}[t!] \centering
\includegraphics[width=0.475\textwidth]{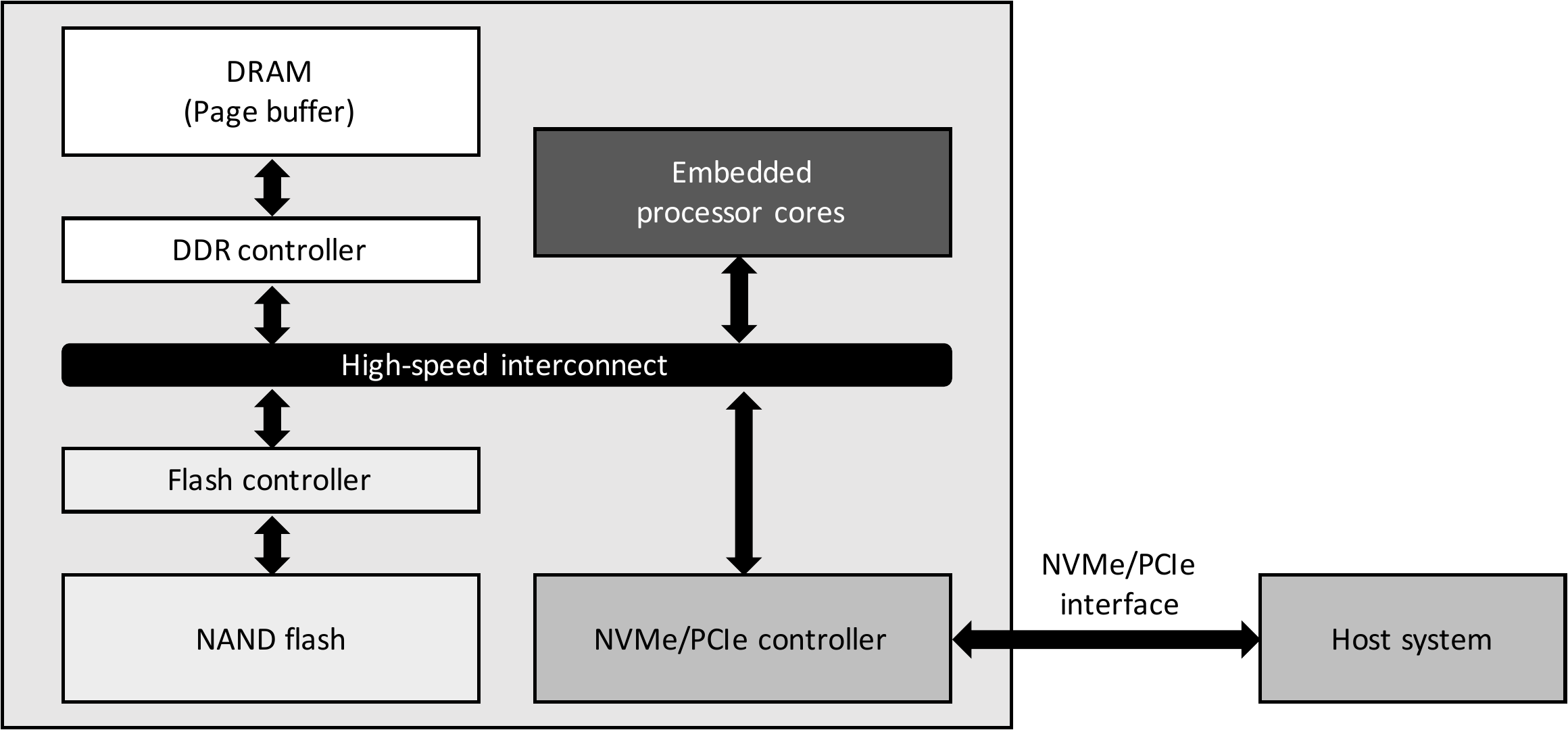}
\caption{
A firmware-based CSD architecture. The embedded processor cores 
	execute routine SSD firmware tasks (e.g., FTL) to handle
	the host-side I/O block requests and
		generate flash page read/write requests from/to the flash devices, buffering them
		inside an 	
on-device DRAM buffer (referred to as \emph{SSD's DRAM page buffer}) to send it back
to the host CPU. 
\proposed utilizes these embedded processor cores to run neighbor sampling directly
off of SSD's DRAM page buffer, seeking to reduce latency.
}
\vspace{-0.8em}
\label{fig:csd_arch}
\end{figure}

{\bf Hardware.} The
	\proposed ISP is implemented as part of the firmware within the 
	SSD (\fig{fig:csd_arch}). Such design decision allows \proposed  to
	maintain compatibility with existing hardware (SSD hardware) and software
	(the OS and the NVMe protocol) stack. By initiating neighbor sampling near
	SSD, however, \proposed is able to increase 
	\emph{effective} throughput for subgraph generation by utilizing the internal SSD
	bandwidth. Because the sampled, subgraphs
	are densely packed within the SSD$\rightarrow$DRAM returned
	logical blocks, \proposed can significantly reduce unused data transferred
	over PCIe, achieving high neighbor sampling throughput.

{\bf Software.} \proposed is  designed to reduce the command and
control overheads in the OS and host driver stack by optimizing the software
runtime for latency rather than locality.  Instead of needlessly incurring
several tens of microseconds of latency to maintain the opportunistic OS page
cache, \proposed allocates a user-space scratchpad buffer to \emph{manually}
orchestrate high locality data movements using Linux direct I/O.  This allows \proposed
software runtime system to \emph{bypass} the OS page cache thereby
significantly reducing the latency overheads of traversing through the system
software stack. Additionally, \proposed host driver employs an ISP instruction
that coalesces multiple I/O commands under a single NVMe command, further
reducing latency.
Such lightweight
communication path is utilized by our \proposed ISP architecture to 
streamline the subgraph generation process with high performance.

\subsection{Hardware Acceleration using In-Storage Processing Architectures}
\label{sect:proposed_hw}

{\bf Why choose firmware-based (and not FPGA-based) CSDs for ISP?} \proposed
ISP unit accelerates the subgraph generation process by offloading the neighbor
sampling operator as part of SSD's firmware execution. There are currently two
prominent approaches in designing a computational storage device (CSD) with ISP
capabilities. One popular approach in designing CSDs is to utilize the embedded
cores within the SSD device for in-storage processing at the SSD's
firmware level (e.g., OpenSSD~\cite{openssd}, NGD system's
		Newport~\cite{newport,newport_product_webpage}, Biscuit~\cite{biscuit}).  An alternative
CSD design point is to utilize FPGA circuitry integrated near the SSD device
and conduct ISP at the hardware level (e.g., Samsung's
		SmartSSD~\cite{smartSSD}, Eideticom's NoLoad CSP~\cite{NoLoadCSP}).
In the remainder of this paper, we refer to each of these CSD types as
\emph{firmware-based} CSD and \emph{FPGA-based} CSD, respectively.

\begin{figure}[t!] 
\centering
\includegraphics[width=0.19\textwidth]{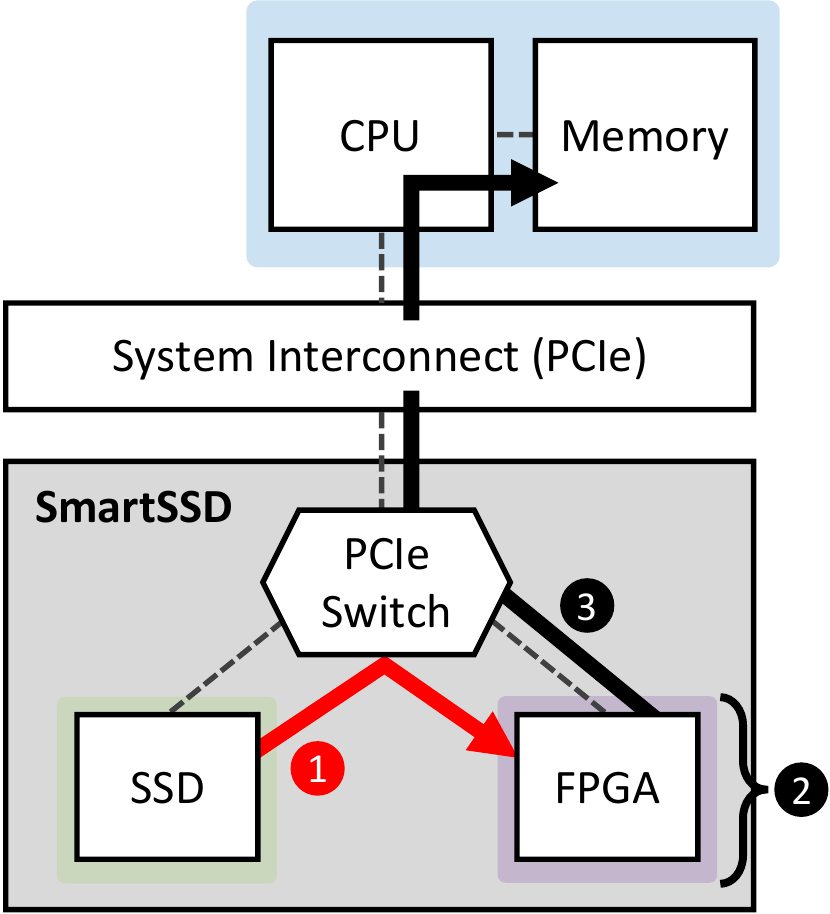}
\caption{ 
Key steps undertaken when conducting neighbor sampling over an FPGA-based CSD (e.g., SmartSSD~\cite{smartSSD}).
Because the SSD and FPGA communicates over a PCIe switch integrated within the CSD device, neighbor sampling over an FPGA-based
	CSD incurs
a two-step P2P transfer (i.e., SSD$\rightarrow$FPGA and then FPGA$\rightarrow$CPU), unlike a firmware-based CSD where
a single SSD$\rightarrow$CPU data transfer is invoked.
}   
\label{fig:csd_comparison}
\vspace{-0.5em}
\end{figure}

We observe that offloading neighbor sampling to an FPGA-based CSD is not
cost-effective because of the following two factors.  First, as our
characterization study revealed in \sect{sect:characterization}, the neighbor
sampling operator is mostly dominated by random data lookups with very low
compute intensity. Therefore, synthesizing hardwired (FPGA) circuitry to merely
conduct data gathers and scatters can 
underutilize the FPGA reconfigurable logic (although this problem can potentially be alleviated by spatially sharing
		the FPGA among different processes). But more crucially, neighbor
sampling offloaded to an FPGA-based CSD incurs a two-step P2P data transfer
which adds significant performance overhead:
1) SSD$\rightarrow$FPGA to conduct the neighbor sampling using FPGA and
generate the subgraph (step $1$/$2$ in \fig{fig:csd_comparison}), and 2) FPGA$\rightarrow$CPU to transfer the subgraph to host CPU
memory (step $3$ in \fig{fig:csd_comparison}). We
observe that the latency overheads of such two-step neighbor sampling outweighs
the benefits of ISP acceleration, providing little benefits over the baseline
mmap-based SSD system.

To this end, \proposed employs a firmware-based CSD substrate to implement our
ISP architecture (\fig{fig:csd_arch}). Because neighbor sampling is mostly composed of irregular data
		lookups with low compute intensity, it suits well to the wimpy embedded
		cores (relative to the host-side x86 CPUs)  equipped within firmware-based
		CSDs.  Later in \sect{sect:eval_vs_smartssd}, we quantitatively
		evaluate both CSD design points for the completeness of our study.

\begin{figure}[t!] 
\centering
\vspace{-1em}
\subfloat[]{
	\includegraphics[width=0.47\textwidth]{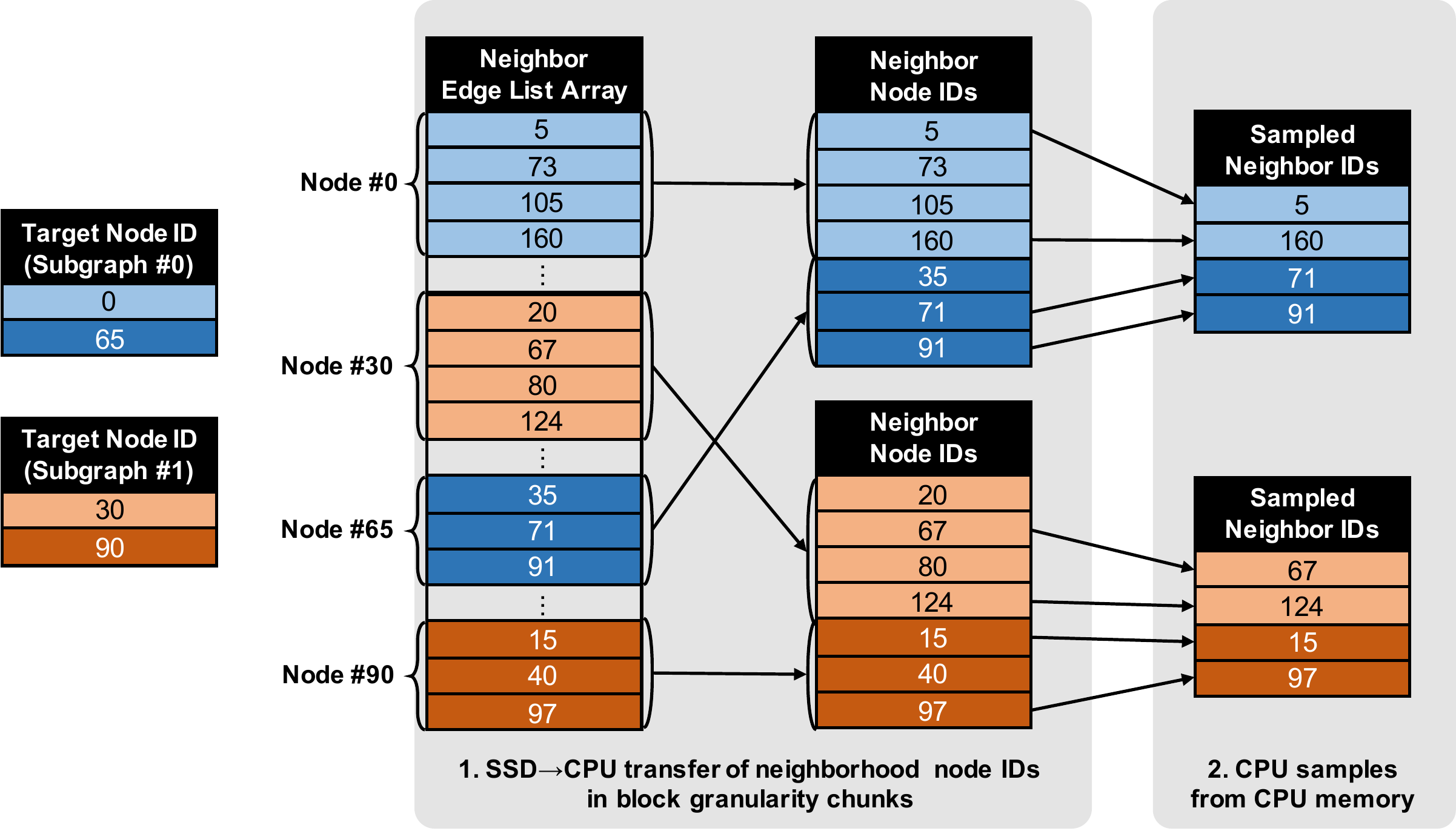}
}
\vspace{0.5em}
\subfloat[]{
	\includegraphics[width=0.47\textwidth]{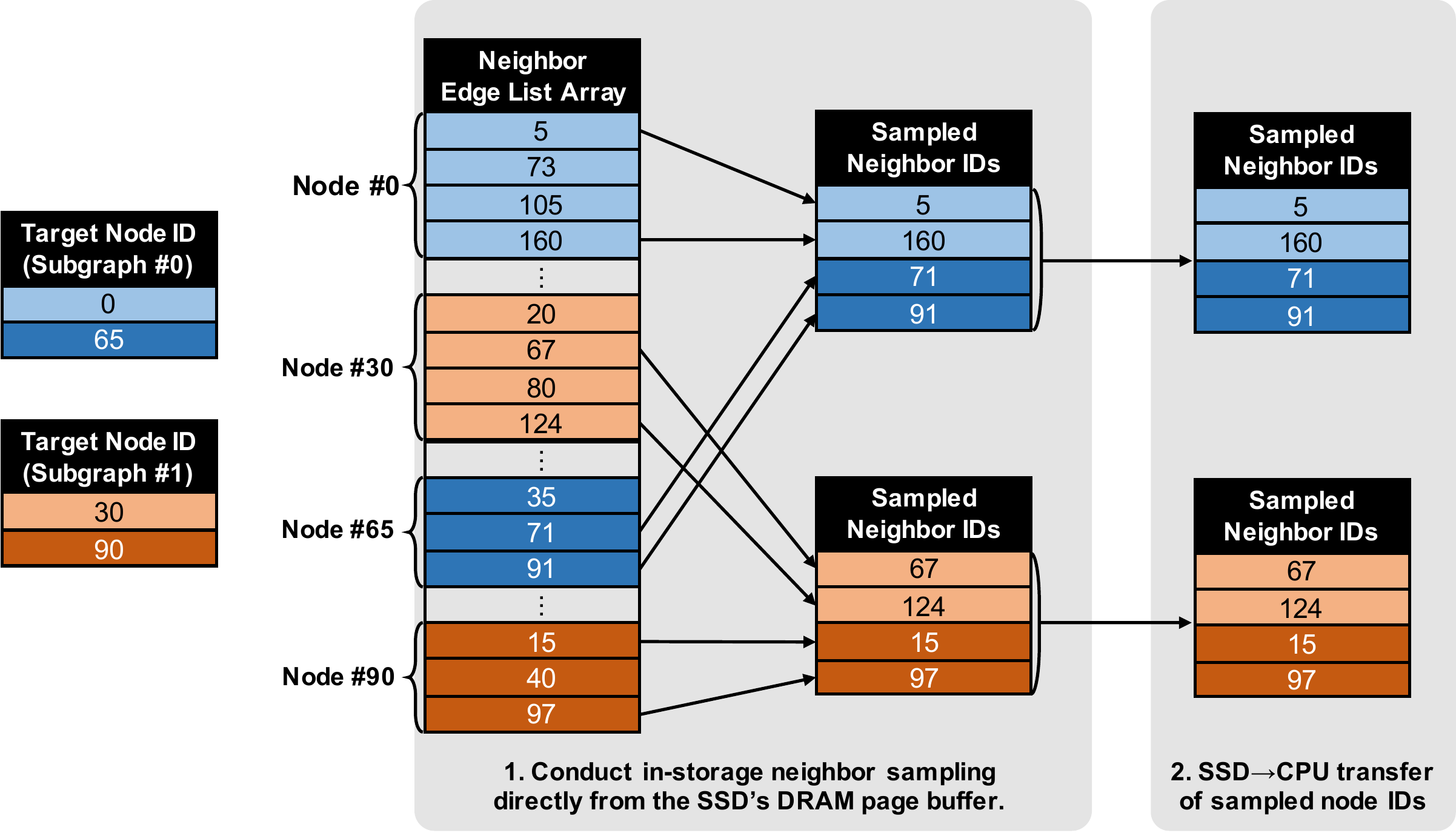}
} 
\vspace{0.1em} 
\caption{
(a) Subgraph generation under the baseline SSD-centric system and (b)	our \proposed ISP based neighbor sampling operation, amplifying effective throughput in generating the subgraph. The example illustrates the process of extracting two subgraphs over two separate mini-batches (blue: mini-batch \#0, orange: mini-batch \#1), each generating a subgraph from two target nodes (target node ID of (blue: $0$/$65$) and (orange: $30$/$90$) and its neighborhood nodes (sampling rate: $2$ neighbors).
}
\label{fig:sampling}
\vspace{-0.5em}
\end{figure}

{\bf Key observations and proposed approach.} \fig{fig:sampling} illustrates the key
intuition behind \proposed ISP neighbor sampling operator.  Here,
					the neighbor edge list array stores all the neighborhood nodes' IDs around a
					given graph node in a sequential manner, for \emph{all} the nodes within the
					graph dataset.
					Under the baseline mmap-based SSD, the CPU
					initiates a memory-mapped block I/O read requests to this data structure as means
					to fetch all the target node's neighborhood ID list into main memory. Because the
					target node IDs within a mini-batch are rarely aligned consecutively,
					there exists little spatial locality that can be reaped out by the CPU
					during neighbor sampling. Consequently, the baseline
					SSD-centric system ends up generating a large number of I/O block
					fetch requests, proportional to the number of target nodes subject
					for neighbor sampling (e.g., one edge list ``chunk'' is fetched
							per each target node in \fig{fig:sampling}(a)).  This results in
					a significant overfetching of useless data from SSD$\rightarrow$CPU, leading to severe
					underutilization of precious I/O bandwidth.  \proposed ISP
					architecture, on the other hand, performs fine-grained gather
					operations from the edge list array \emph{directly} from the SSD's
					DRAM page buffer, constructing a dense ID list of ``sampled'' neighborhood
					node IDs (\fig{fig:sampling}(b)). Because this dense,  sampled node
					ID list (i.e., the subgraph itself) is subject for
					SSD$\rightarrow$CPU transfer, \proposed is able to significantly
					amplify the \emph{effective} throughput of neighbor sampling.

\begin{figure}[t!] \centering
\includegraphics[width=0.47\textwidth]{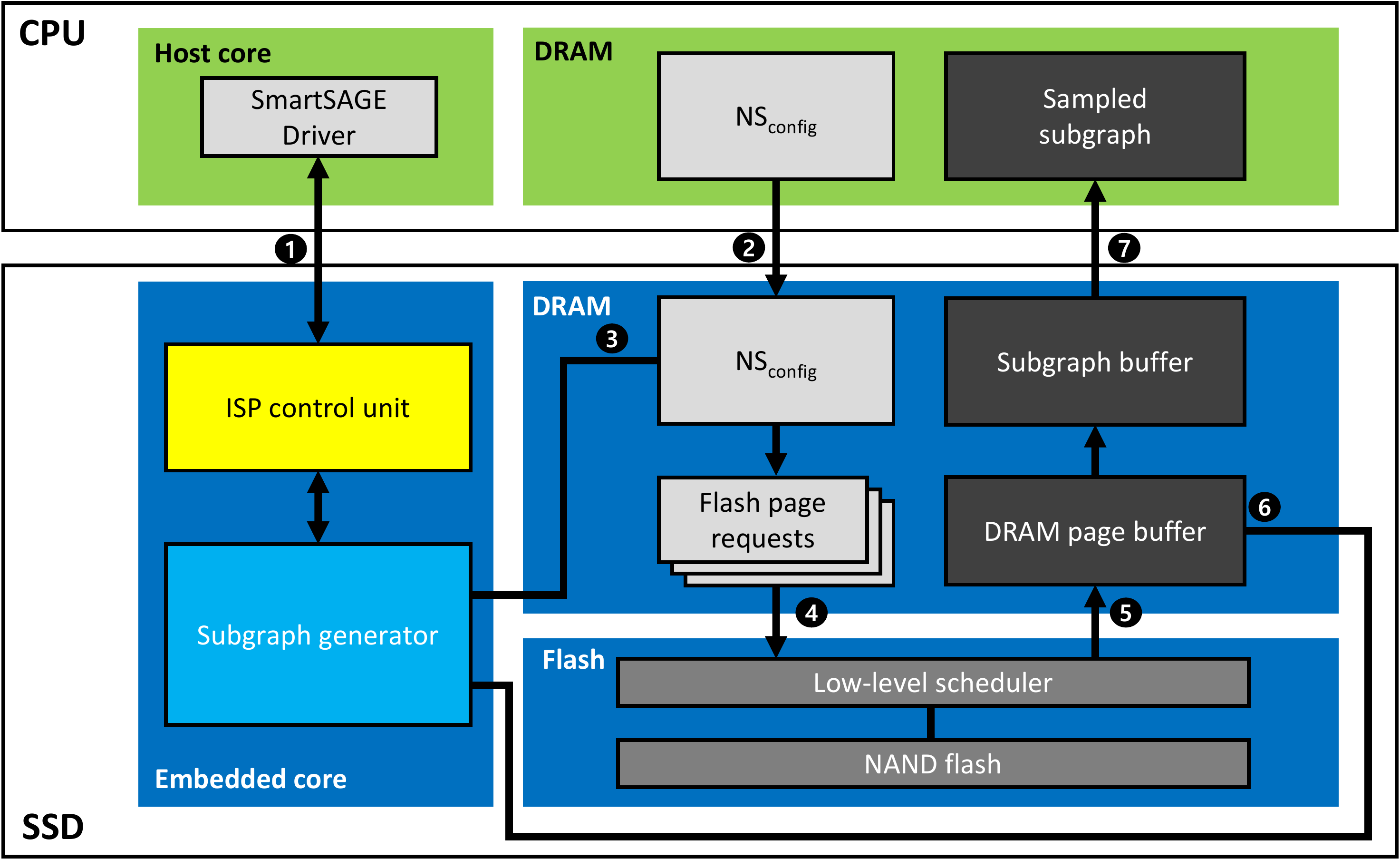}
\caption{
	Major components of SmartSAGE ISP architecture and the key steps undertaken during the lifetime of in-storage neighbor sampling.
}
\vspace{-0.5em}
\label{fig:isp_arch_and_dataflow}
\end{figure}

{\bf Hardware/software interaction in ISP neighbor sampling.} We now detail the key
components of \proposed  ISP design
as well as the associated hardware/software interactions involved during in-storage neighbor sampling (\fig{fig:isp_arch_and_dataflow}).
\proposed adds two major components within the SSD firmware: 1) The ISP
	control unit handles the CPU$\rightarrow$SSD offloaded subgraph generation request,
					generating the necessary set of commands to send to the subgraph generator.
						2) The subgraph generator is in charge of extracting out the subgraph from the
						large-scale input graph, sending series of flash page read
						requests to the low-level flash devices. Once the requested flash pages are returned
						and subsequently cached into
						SSD's DRAM page buffer, the subgraph generator initiates in-storage neighbor sampling to
						gather the sampled nodes of the subgraph under construction. Below we
						summarize the major steps undertaken during \proposed's subgraph
						generation.

\begin{enumerate} \item Our custom designed \proposed driver (detailed in
		\sect{sect:proposed_sw}) sends a subgraph generation request to the SSD
firmware intitially as an NVMe write command, which includes a pointer to the
\emph{neighbor sampling configuration data} (\nsconfig) stored inside CPU memory. \nsconfig 
contains key parameters of the
sampling operation, e.g., number of target nodes as well as their logical block
address, neighborhood node IDs to sample, and
other metadata (step \ding{182} in \fig{fig:isp_arch_and_dataflow}).

\item When the SSD firmware receives the subgraph generation request, it
triggers a DMA access to CPU memory to copy the \nsconfig data from
CPU$\rightarrow$SSD using SSD's NVMe host controller (step \ding{183}). Once
the SSD receives the \nsconfig data, it goes through several stages
to prepare for subgraph generation, one major step being the address
translation process (from the logical address to flash's physical page address)
to determine where within the flash devices should the subgraph generator send
flash page read requests to (step \ding{184}). A given target node's neighbor
nodes' ID list can potentially require multiple flash page read requests
depending on the number of edges connected to the target node.

\item The flash page read requests are sent to the pending flash page request queue to 
prepare for in-storage neighbor sampling,
	which the low-level flash controller utilizes to kick off flash page read operations
(step \ding{185}). 

\item Once a flash page read request is serviced, the corresponding neighbor
edge list will be cached inside SSD's DRAM page buffer (step \ding{186}).
\proposed utilizes the SSD's embedded cores to conduct fine-grained neighbor
sampling over the neighbor edge list, stored inside the SSD's DRAM
page buffer.  The sampled nodes are collected into ISP's pending subgraph
buffer (step \ding{187}).

\item Once all the target nodes' neighbor IDs are sampled, the sampled subgraph is ready to be transferred
back to the CPU. At the high level of the SSD firmware polling loop, our scheduler checks for completed
subgraph generation requests. If the sampled subgraph is ready and the NVMe host controller is available,
we initiate a				 
SSD$\rightarrow$CPU DMA write operation to copy back the subgraph information
into CPU DRAM (step \ding{188}).

\end{enumerate}

\subsection{Latency-optimized Runtime and Host Driver Stack}
\label{sect:proposed_sw}

Although our ISP design helps significantly reduce the latency of
neighbor sampling, there are still significant performance improvement
opportunities to be reaped out. As discussed in \sect{sect:characterization}, 
the baseline mmap-based SSD experiences significant slowdown as the
locality-optimized OS page cache is rarely useful in reducing I/O access
time.   Additionally, recall from \fig{fig:sampling} that the baseline
SSD invokes a large number of I/O block fetch requests per each
mini-batch training as the data accesses to the neighbor edge list exhibit
low spatial locality (i.e., only a single edge list ``chunk'' can be requested over 
a single I/O block fetch request), which adds another layer of latency
penalty. This is because servicing any given I/O block fetch request involves
context switching through the user$\leftrightarrow$kernel space when 
traversing through the host driver stack.

We  design our software runtime and the host driver stack
to be optimized for latency first and locality second, as detailed below.

\begin{figure}[t!] \centering
\includegraphics[width=0.475\textwidth]{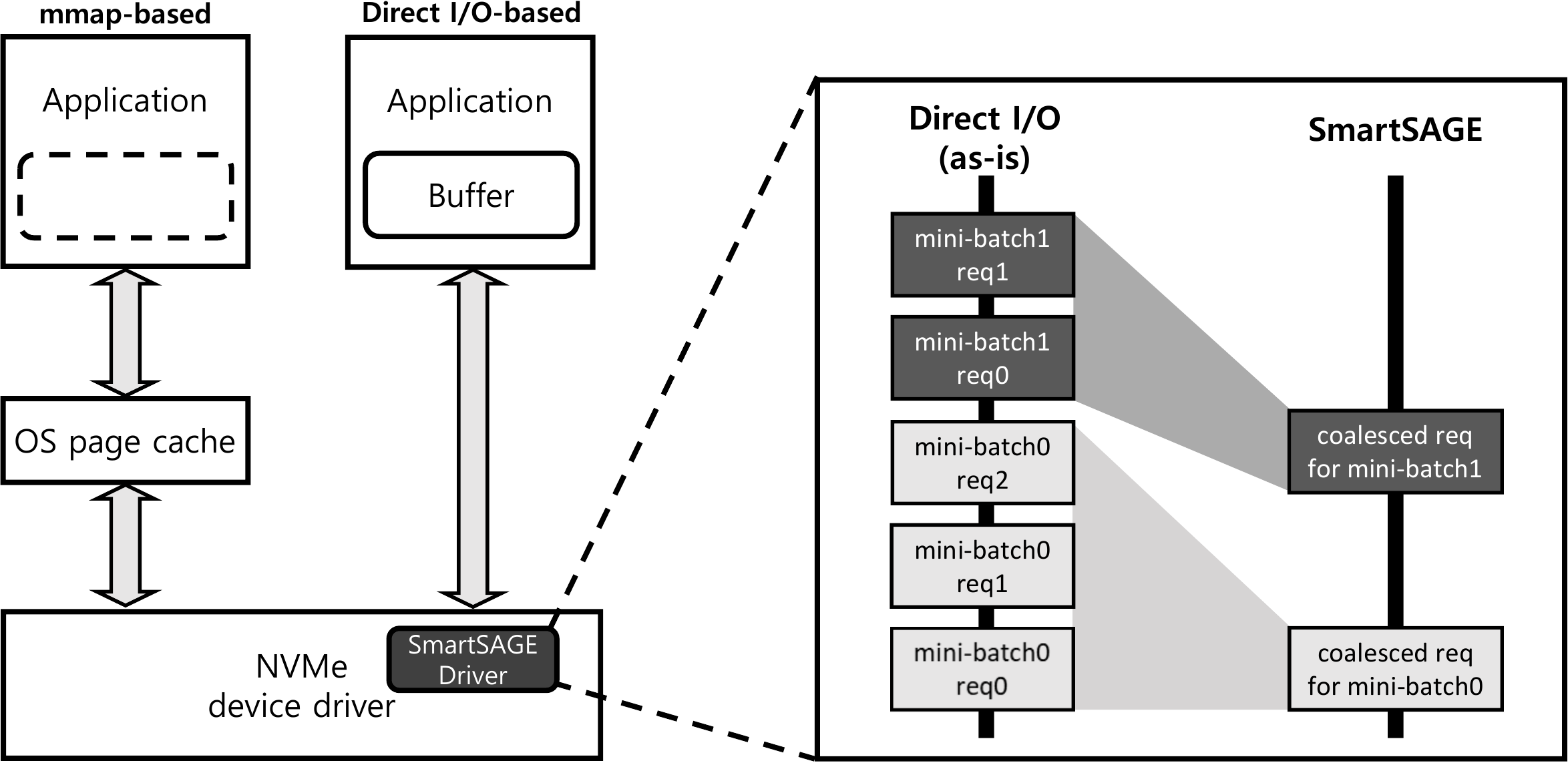}
\caption{
\proposed's latency-optimized direct I/O based software runtime (left) and host driver stack employing I/O command coalescing (right).
}
\vspace{-1.3em}
\label{fig:proposed_runtime}
\end{figure}

{\bf Direct I/O.} Linux comes with a direct I/O feature where the file
read/write operations go directly from the user-space application to the
storage device, \emph{bypassing} the OS page cache completely
(\fig{fig:proposed_runtime}). Given the locality limited nature of neighbor
sampling, our runtime system utilizes such feature (\texttt{O\_DIRECT} flag) to
allocate a user-space buffer to \emph{manually} orchestrate high locality data
movements between SSD$\leftrightarrow$CPU, without relying on the opportunistic
OS page cache. As we quantitatively analyze in \sect{sect:eval_multiple_worker}, our direct I/O
based software runtime alone, even without the ISP feature, helps reduce
neighbor sampling latency by $2.9\times$ compared to the baseline
mmap-based SSD system.

{\bf I/O command coalescing.} As detailed in \sect{sect:proposed_hw}, the
in-storage neighbor sampling operation is invoked by sending a subgraph
generation request in the form of an NVMe write command to the SSD. Unlike the
baseline SSD-centric system which spawns off multiple high latency I/O fetch
requests for a single subgraph generation, \proposed's host driver encapsulates
the \emph{entire} gather/scatter operation of neighbor sampling under a
\emph{single} NVMe transaction.  More concretely, all the key parameters of
subgraph generation (e.g., logical block address of all target nodes,
		neighborhood node ID lists to sample, etc) are stored as part of the
\nsconfig data (\fig{fig:isp_arch_and_dataflow}), which the SSD
firmware retrieves through a one-time CPU$\rightarrow$SSD DMA transaction.
This allows \proposed to significantly reduce the number of I/O commands
invoked to generate a single subgraph, reducing command and control overheads
of the host driver stack and further reducing latency.

{\bf NVMe compatibility and system integration.} Our custom I/O interface is
implemented using the	\texttt{ioctl()} system call, which maintains complete
compatibility with current NVMe protocols. The subgraph generation request
is indicated using a single unused command bit, which the SSD firmware utilizes
to invoke in-storage neighbor sampling. Aside from this special-purpose bit, the
hardware/software interface utilizes the exact same command structure of
conventional I/O read/write commands and the CPU$\rightarrow$SSD copy of
\nsconfig data as well as SSD$\rightarrow$CPU copy of the final, neighbor sampled subgraph is
orchestrated using existing DMA engines.

Because \proposed's runtime and host driver stack maintains compatibility
with current NVMe protocol and the ISP operator is implemented as part of
the SSD firmware, \proposed is fully compatible with existing CSD architectures.

%% file: tex/methodology.tex
\section{Methodology}
\label{sect:methodology}

\begin{table}[t!]
\centering
\caption{Graph dataset information.}
\resizebox{\columnwidth}{!}{
\begin{tabular}{l cccc cccc}
\toprule
& \multicolumn{3}{c}{In-memory} & \multicolumn{3}{c}{Large-scale} & \\
\cmidrule(lr){2-4} \cmidrule(lr){5-7}
Dataset & Nodes   & Edges & Size (GB) & Nodes  & Edges & Size (GB) & Features \\
\midrule
Reddit    & 233.0K & 114.6M & 0.8 & 37.3M  & 53.9B & 402 & 602 \\
Movielens  & 5.5M & 6.0B & 45 & 22.2M  & 59.2B & 442 & 1K \\
Amazon   & 42.5M & 1.3B & 9.7 & 265.9M  & 9.5B & 75 & 32  \\
OGBN-100M   & 89.6M & 3.2B & 26 & 179.1M  & 5.0B & 41 & 32  \\
Protein-PI  & 907.0K & 317.5M & 2.4 & 9.1M  & 8.8B & 66 & 512  \\
\bottomrule
\end{tabular}
}
\label{tab:dataset}
\vspace{-.5em}
\end{table}

\textbf{Hardware/software platform.} We use PyTorch Geometric to implement a
GraphSAGE based GNN training pipeline. When evaluating end-to-end
training performance, we employ a CPU-GPU system containing an Intel
Xeon Gold 6242 CPU with $192$ GB of DRAM and NVIDIA's Tesla T4 GPU.

\textbf{CSD platform.} We implement our neighbor sampling ISP operator
using the open-source Cosmos+ OpenSSD~\cite{openssd}. OpenSSD
contains a fully functional NVMe flash SSD with $2$ TB of storage and a
customizable SSD firmware executed by a dual core ARMv7 Cortex-A9 processor.
The host interface controller of OpenSSD communicates with the CPU over an
$8$-lane PCIe (gen2) channel. 
We design \proposed's ISP operator within
OpenSSD's firmware and measure wall clock time for evaluating performance.

{\bf Comparison to alternative training systems.} While we demonstrate \proposed's
merits using OpenSSD, for the completeness of our
study, we also evaluate \proposed's ISP neighbor sampling operator on top of an
FPGA-based CSD using Samsung-Xilinx's SmartSSD~\cite{z_ssd} (\sect{sect:eval_vs_smartssd}).
We also compare \proposed against a training system utilizing Intel's Optane DC
Persistent Memory Module (PMEM) that is installed on the memory bus (NVDIMM),
	providing $768$ GB of capacity  to store the graph datasets
	(\sect{sect:eval_end_to_end}).
	 
\begin{figure}[t!] \centering
\vspace{-1em}
\subfloat[Reddit]{
\includegraphics[width=0.5\textwidth]{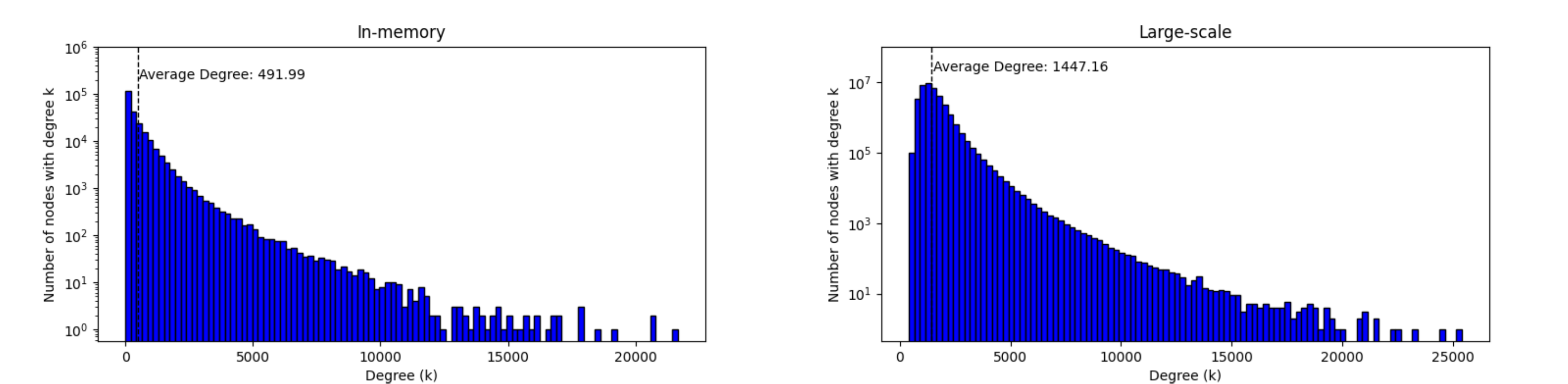}
	\label{fig:degree_distribution_reddit}
}
\vspace{0.1em}
\subfloat[Protein-PI]{
\includegraphics[width=0.5\textwidth]{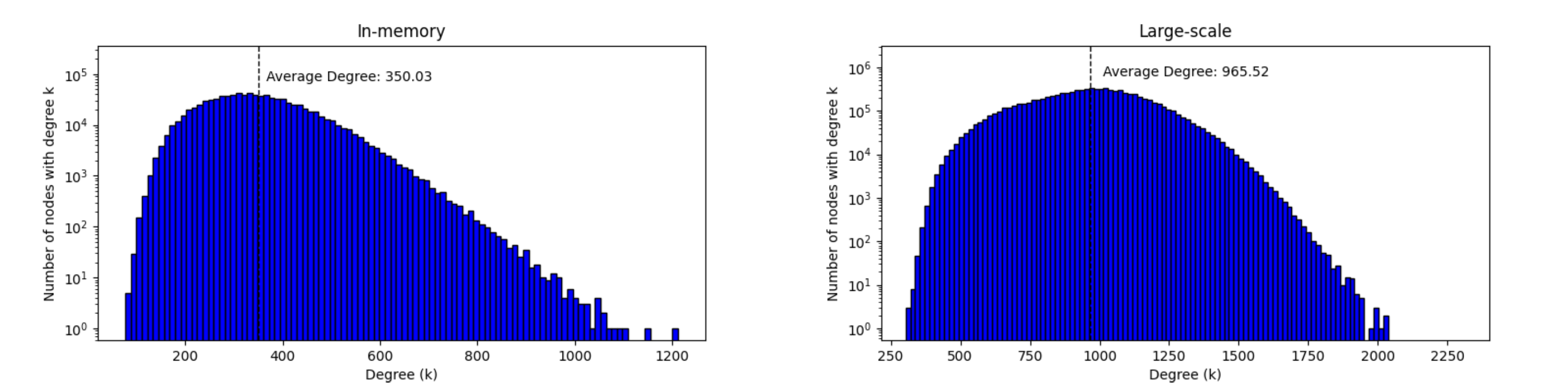}
\label{fig:degree_distribution_ppi}
}
\caption{
Degree distribution of the in-memory (left) vs. large-scale (right) graph datasets for a subset of our datasets.  As depicted, despite
		Kronecker fractal expansion enabling the number of nodes as well as edges
		to increase dramatically (represented by the increase in the number of nodes
				with degree-K in the y-axis), the overall power-law distribution of
		each dataset, before/after fractal expansion is applied remains similar.
}
\vspace{-.5em}
\label{fig:expanded_graph}
\end{figure}

\textbf{Dataset.} Existing graph datasets utilized for GNN studies are
generally at small-scale that comfortably fit within main memory, defeating the
main purpose of our study. Prior work on Kronecker Graph
Theory~\cite{fractal_expansion_mlperf} suggests graph fractal expansion methods
to expand the scale of the input graph dataset while properly maintaining the
graph's  distinctive characteristics (e.g., power-law degree distribution,
		community structure, etc). We therefore employ Kronecker fractal expansion
method~\cite{fractal_expansion_mlperf} to synthetically generate large-scale
graph datasets, the key features of which are summarized in \tab{tab:dataset}
(the ``\emph{Large-scale}'' column). Prior work~\cite{densification}
observed that real-world graphs that become larger with more nodes and edges
generally exhibit higher average degrees (i.e., the rate at which the number of
		edges increase is faster than the increase in number of nodes), a property
known as the \emph{densification power law}. The synthetically generated
large-scale datasets are designed to properly reflect such behavior,
	represented by the higher average degree of large-scale datasets vs.
	in-memory datasets (\fig{fig:expanded_graph}). The graph datasets are all compressed in CSR (compressed sparse row)
	format and we employ the default configuration of GraphSAGE when
	training our GNN algorithm with a mini-batch size of $1024$.
In \sect{sect:eval_sensitivity}, we discuss the sensitivity of \proposed when deviating from these default configurations.

%% file: tex/result.tex
\section{Evaluation} 
\label{sect:results}

This section first focuses on exploring the following three design points: 1)
the baseline mmap-based SSD training system (\ssd{mmap}), and two design points
employing our hardware/software optimizations, namely 2) direct I/O based SSD
training system \emph{without} ISP (\sage{SW}) and 3) our proposed architecture
with all the proposed optimizations in place (\sage{HW/SW}). Later in
\sect{sect:eval_end_to_end}, we compare \proposed against an upper bound, oracular
in-memory processing design point which assumes the CPU contains \emph{infinite}
DRAM capacity so that all the graph datasets can be stored locally in DRAM. \proposed is also
compared against Intel's Optane DC PMEM in \sect{sect:eval_end_to_end} and an ISP
architecture based on an FPGA-based CSD in \sect{sect:eval_vs_smartssd}.

\subsection{``Single'' Worker's Neighbor Sampling Performance}
\label{sect:eval_single_worker}

\begin{figure}[t!] \centering
\includegraphics[width=0.475\textwidth]{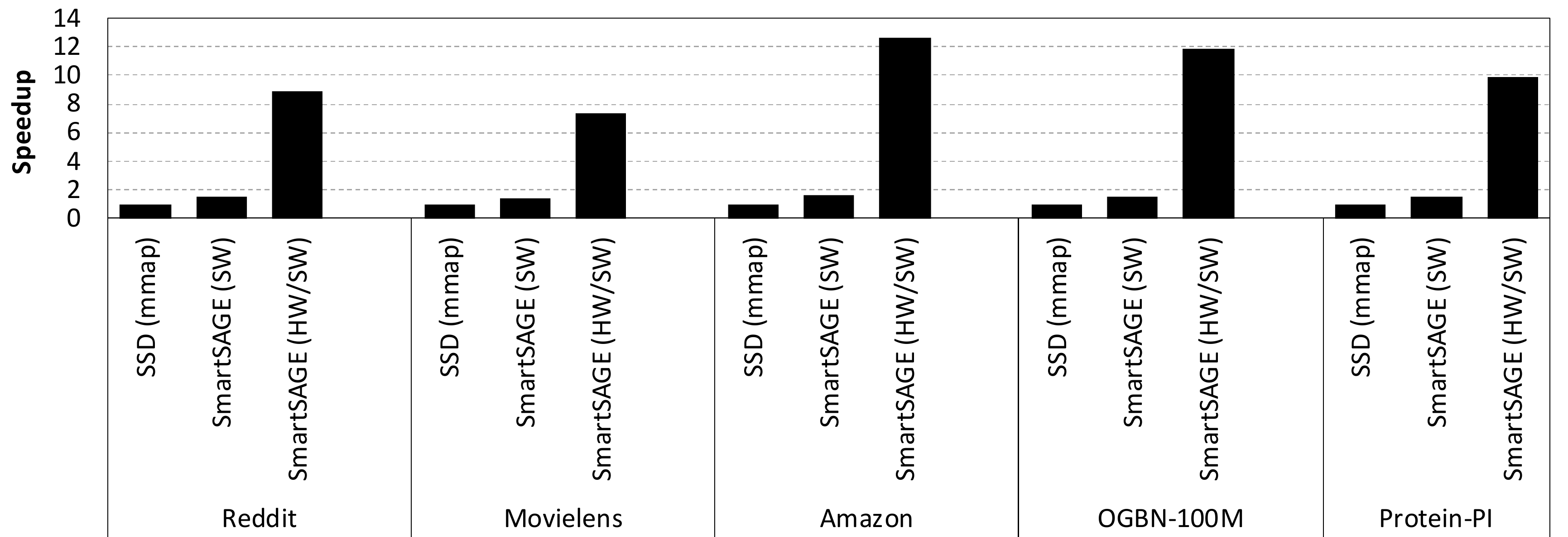}
\caption{
\proposed's speedup for neighbor sampling vs. baseline mmap-based SSD system (single worker). 
}
\vspace{-.5em}
\label{fig:speedup_single_worker}
\end{figure}

 As discussed in \fig{fig:execution_timeline_model}, a GNN training pipeline
 employs a producer-consumer model where multiple CPU-side workers
 independently conduct neighbor sampling for subgraph generation. To precisely
 understand the efficacy of our proposal on improving a 
 CPU-side worker's neighbor sampling operation, we first instantiate just a single
 worker for subgraph generation and evaluate its performance 
(\fig{fig:speedup_single_worker}). Compared to
 the baseline mmap-based SSD system, our software-only \proposed with direct
 I/O runtime (\sage{SW}) alone provides an average $1.5\times$
		speedup in neighbor sampling. Such result demonstrates the advantage
		of using direct I/O to bypass the OS page cache, which helps optimize 
the data preparation stage
for latency, rather than locality.
		
\begin{figure}[t!] \centering
\includegraphics[width=0.475\textwidth]{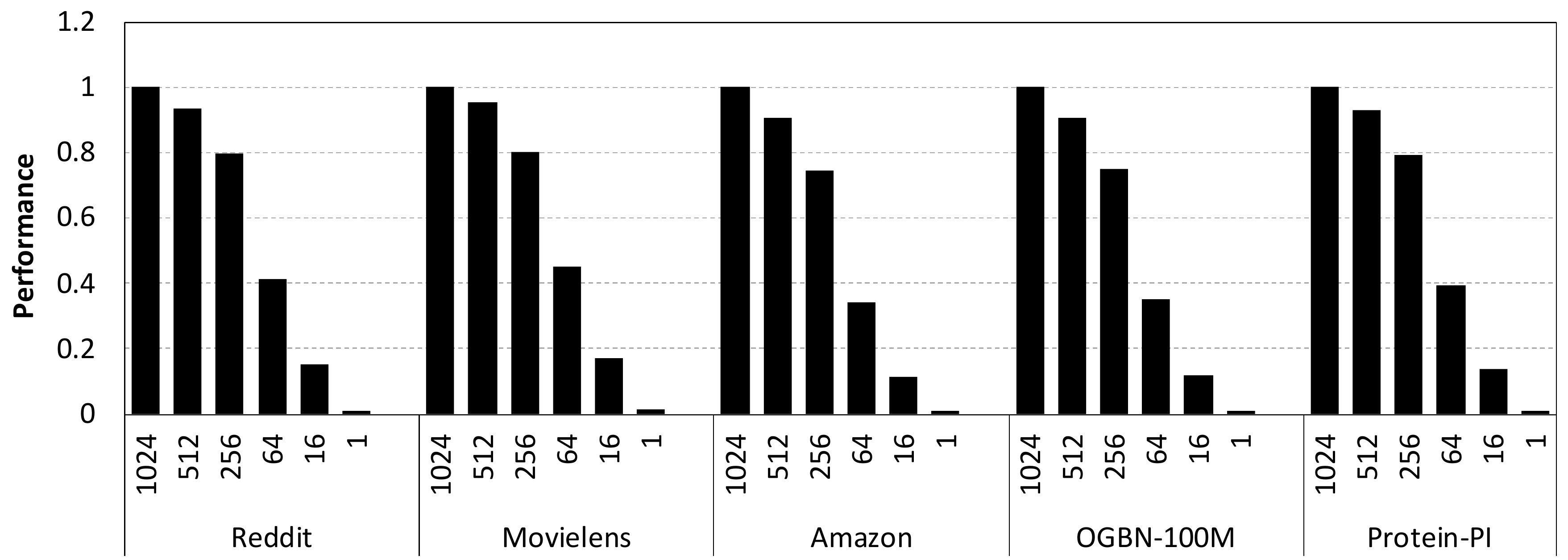}
\caption{
Effect of reducing the I/O command coalescing granularity on \sage{HW/SW} speedup.
}
\vspace{-1.2em}
\label{fig:sensitivity_io_coalescing}
\end{figure}

		The neighbor sampling performance is enhanced even further with \proposed's
		ISP architecture (\sage{HW/SW}), providing an additional average speedup of
		$6.6\times$ (maximum $7.8\times$) over the software-only
		\sage{SW}. Overall, \sage{HW/SW} achieves an average $10.1\times$
		(maximum $12.6\times$) speedup vs. the mmap-based SSD system, successfully
		resolving the bottlenecks of the baseline architecture.

		It is worth mentioning that the significant speedup \proposed achieves is
		not just a result of the hardware-level acceleration of neighbor sampling
		or direct I/O based software runtime; also equally important is our
		co-designed host driver stack that minimizes the I/O command and control
		overheads.  In
		\fig{fig:sensitivity_io_coalescing}, we show the effect of reducing the I/O
		command coalescing granularity on \sage{HW/SW}'s speedup.  The default
		configuration of  \sage{HW/SW} coalesces \emph{all} the target nodes'
neighbor sampling within a given mini-batch (i.e., $1024$ target nodes, the
		leftmost design point) under a \emph{single} NVMe command, which is
encapsulated inside a single \nsconfig (neighbor sampling configuration data,
		see \fig{fig:isp_arch_and_dataflow}). As the coalescing granularity becomes
smaller (from left to right in the x-axis), the latency overheads of sending
the I/O commands to the SSD start outweighing the benefits provided with ISP,
		experiencing a significant performance hit.

Overall, the evaluation in this section highlights the effectiveness of
		\proposed's software/hardware co-design; 1) direct I/O, 2) I/O command coalescing,
		and 3) ISP acceleration.

\begin{figure}[t!] \centering
\includegraphics[width=0.475\textwidth]{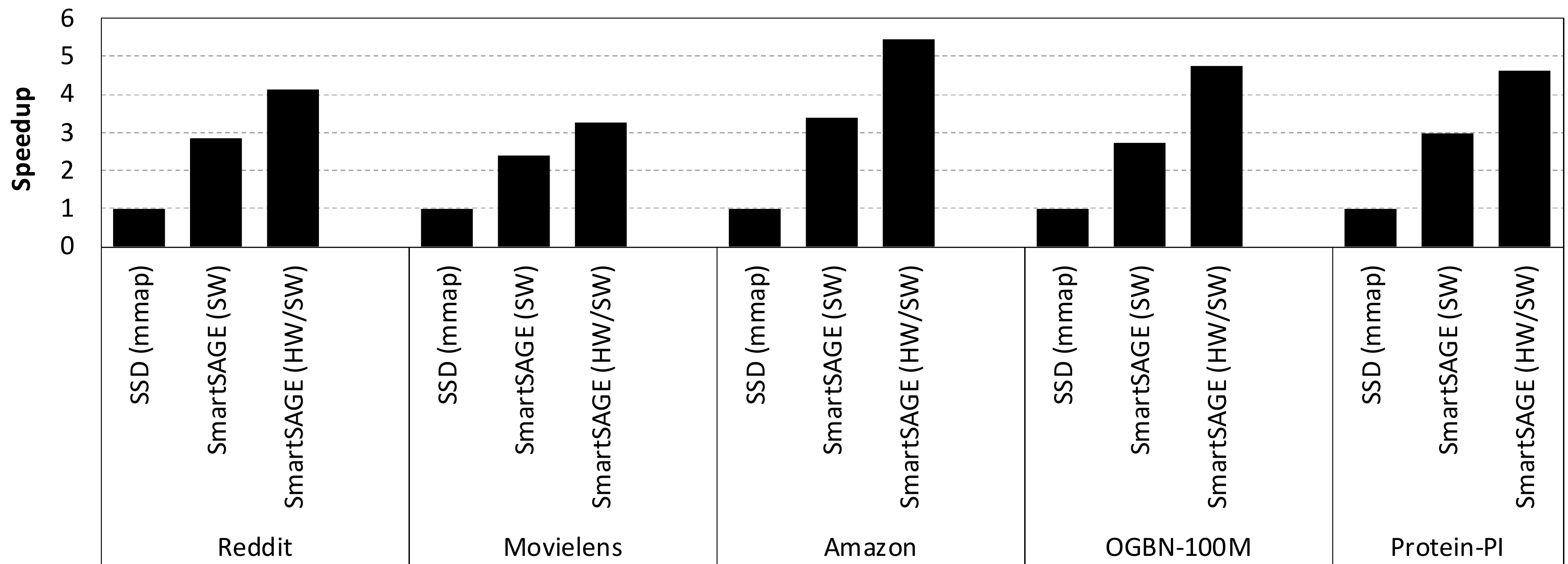}
\caption{
\proposed's neighbor sampling speedup vs. baseline \ssd{mmap} (multiple workers).
Results assume $12$ concurrent workers as performance is at its highest with $12$ workers, for both baseline and \proposed. 
}
\vspace{-1.2em}
\label{fig:speedup_multiple_worker}
\end{figure}

\begin{figure}[t!] \centering
\includegraphics[width=0.475\textwidth]{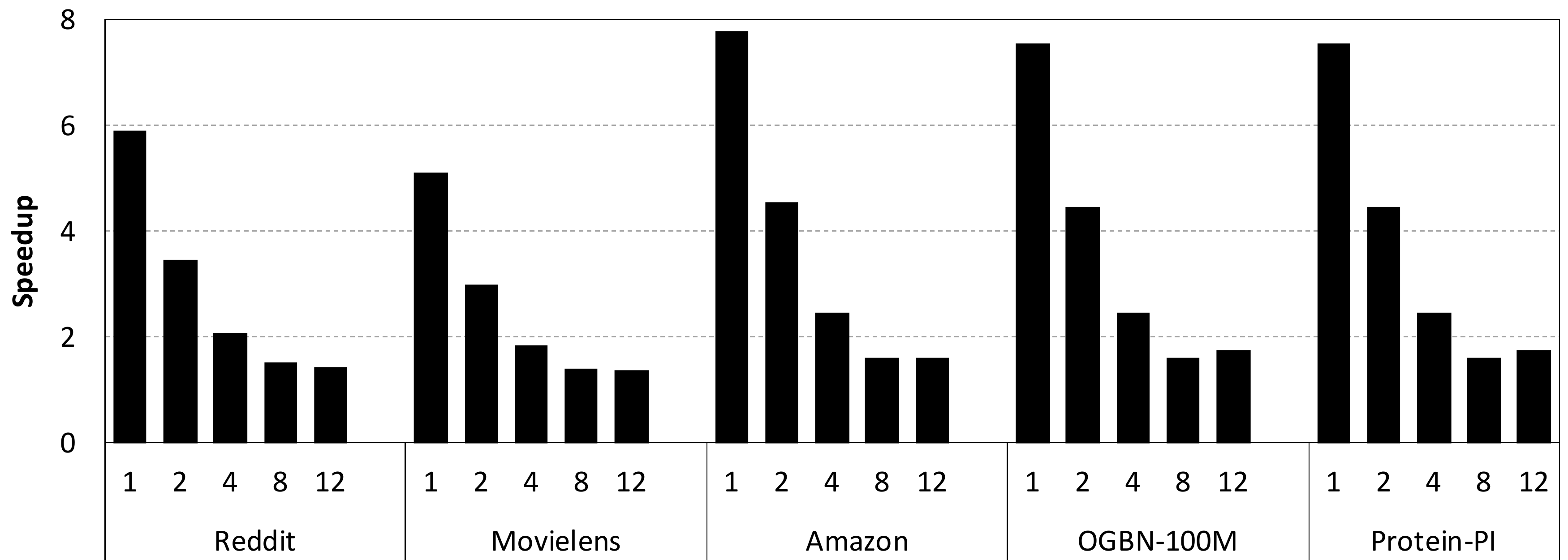}
\caption{
\sage{HW/SW}'s speedup vs. \sage{SW} when scaling up the number of CPU-side workers, from $1$ to $12$.
}
\vspace{-1.2em}
\label{fig:performance_multiworker_scaling}
\end{figure}

\subsection{``Multiple'' Workers' Neighbor Sampling Performance}
\label{sect:eval_multiple_worker}

We now explore the effectiveness of \proposed when multiple workers are
concurrently accessing the storage system for subgraph generation
(\fig{fig:speedup_multiple_worker}).  \sage{HW/SW} provides an average
$4.4\times$ (maximum $5.5\times$) speedup in neighbor sampling
compared to the baseline \ssd{mmap}. Compared to the single worker execution
scenario (\fig{fig:speedup_single_worker}), however, notice that the additional
speedup provided with  \sage{HW/SW} vs.  \sage{SW} is reduced. Our analysis showed
that, when multiple workers simultaneously access the SSD for in-storage
neighbor sampling, OpenSSD's relatively wimpy embedded cores get overwhelmed in delivering sufficient
levels of ISP compute power for high throughput neighbor sampling.
More concretely,
since our OpenSSD based neighbor sampling operator \emph{time-shares} the embedded cores
with the flash management firmware, the interference our neighbor sampling causes
to the SSD firmware  degrades the level of speedup achieved with in-storage neighbor sampling.
Such phenomenon is shown in
\fig{fig:performance_multiworker_scaling}, where \sage{HW/SW}'s speedup vs. \sage{SW}
gets gradually reduced as we add more CPU-side workers.
Nonetheless, we emphasize that our evaluation results are highly
conservative as state-of-the-art CSD
architectures provide far more performance than our OpenSSD platform. 
For
instance, recently announced CSDs like NGD system's
Newport~\cite{newport,newport_product_webpage} provides much faster compute and communication
performance for ISP usage. Unlike our OpenSSD platform which utilizes a dual core ARMv7
Cortex-A9 for firmware execution,
Newport contains four ARMv7 M7 cores for firmware execution with an additional
quad core ARMv8 Cortex-A53 solely dedicated for ISP purposes. 
Due to limited public availability
of Newport CSDs, we employed the
OpenSSD system as a proof of concept to our proposal, demonstrating substantial
improvements in throughput vs. baseline SSD.

In general, we conclude that the \proposed's key insights are directly
applicable to any firmware-based CSD because \proposed maintains full
compatibility with current NVMe protocol and its ISP is
programmed at the software level as part of the SSD firmware.

\begin{figure}[t!] \centering
\includegraphics[width=0.475\textwidth]{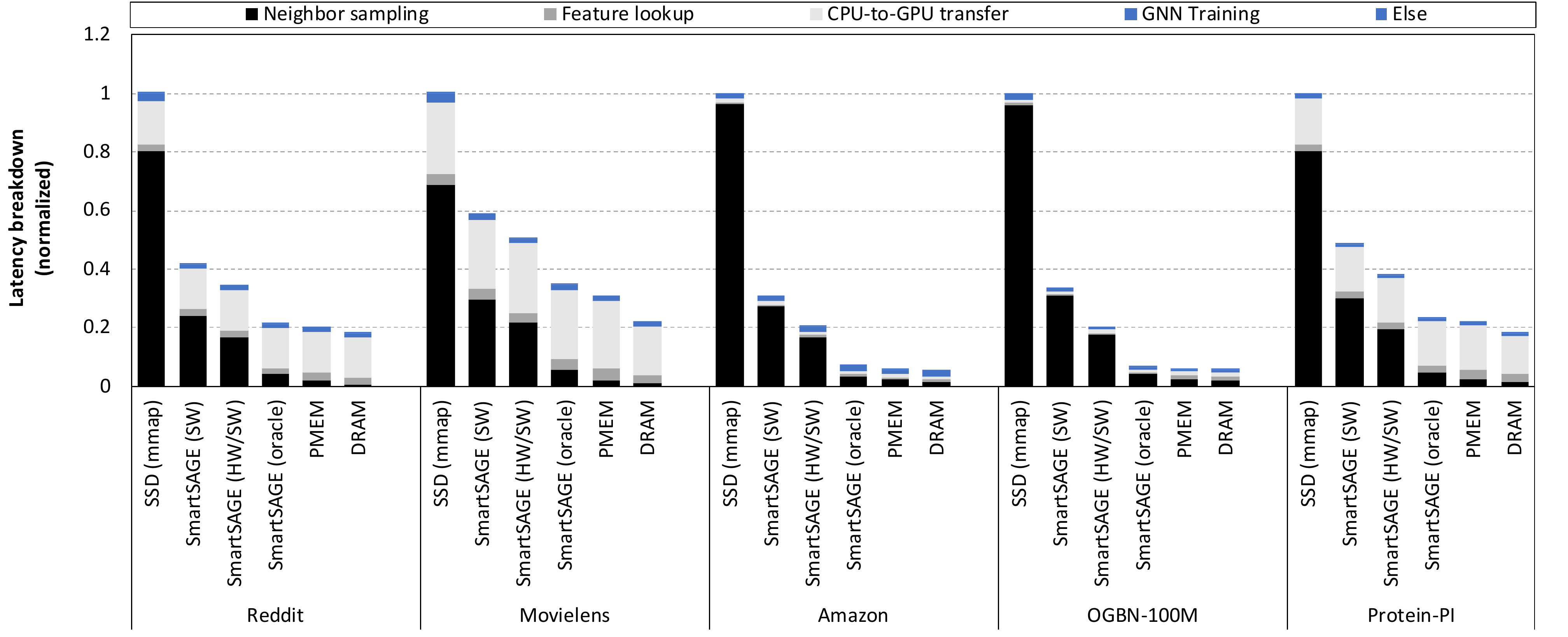}
\caption{
Latency breakdown of end-to-end GNN training time.
}
\vspace{-.5em}
\label{fig:speedup_end_to_end}
\end{figure}

\subsection{End-to-End GNN Training Time}
\label{sect:eval_end_to_end}

\fig{fig:speedup_end_to_end} summarizes the effectiveness of \proposed on
reducing end-to-end GNN training time.  To thoroughly cover the evaluation
space, we explore two in-memory processing based systems: 1) Intel's Optane DC
Persistent Memory Module (PMEM) which can store the entire graph
dataset within its NVDIMMs and 2) an oracular DRAM-only design
which assumes the main memory is infinitely sized to enable large-scale GNN
training at DRAM speed.  We also establish a hypothetical \proposed design
point which assumes that the CSD architecture contains dedicated, ISP-purposed
embedded cores (like Newport CSD~\cite{newport,newport_product_webpage}) such that the opportunties
inherent with our in-storage neighbor sampling can be fully unlocked
(\sage{oracle}). \sage{oracle}'s performance is estimated by assuming the
neighbor sampling speedup achieved under single worker execution
(\fig{fig:speedup_single_worker}) can equally be achieved under multiple
workers. 

Compared to the baseline mmap-based SSD
system, \proposed(HW/SW) provides an average $3.5\times$ (maximum
		$5.0\times$) improvement in training throughput. While such improvements
are impressive, \sage{HW/SW} still incurs an average $60\%$ performance loss
vs. the DRAM-only design point. Nonetheless, such DRAM-only architecture is an unbuildable, upper bound
design point and existing in-memory processing based systems are not able to train the large-scale 
GNNs that our SSD-based system will enable. Intel PMEM does much better than the baseline \ssd{mmap},
		 ``only'' incurring an average $1.2\times$ slowdown vs. the oracular DRAM-only design.
		 However, the performance advantage of PMEM comes at the cost of lower storage density
		and lower GB/\$ vs. SSD. Compared to these high cost in-memory processing based systems,
		\sage{oracle} performs very competitively, achieving an average $70\%$ and $90\%$ of the performance
	of	the DRAM-only and PMEM based systems, respectively.
These results demonstrate that, with newer/future CSD architectures
provisioned with more ISP compute power and faster flash devices, an NVMe SSD based
system can become a viable option for large-scale GNN training while not compromising
on performance.

\begin{figure}[t!] \centering
\includegraphics[width=0.475\textwidth]{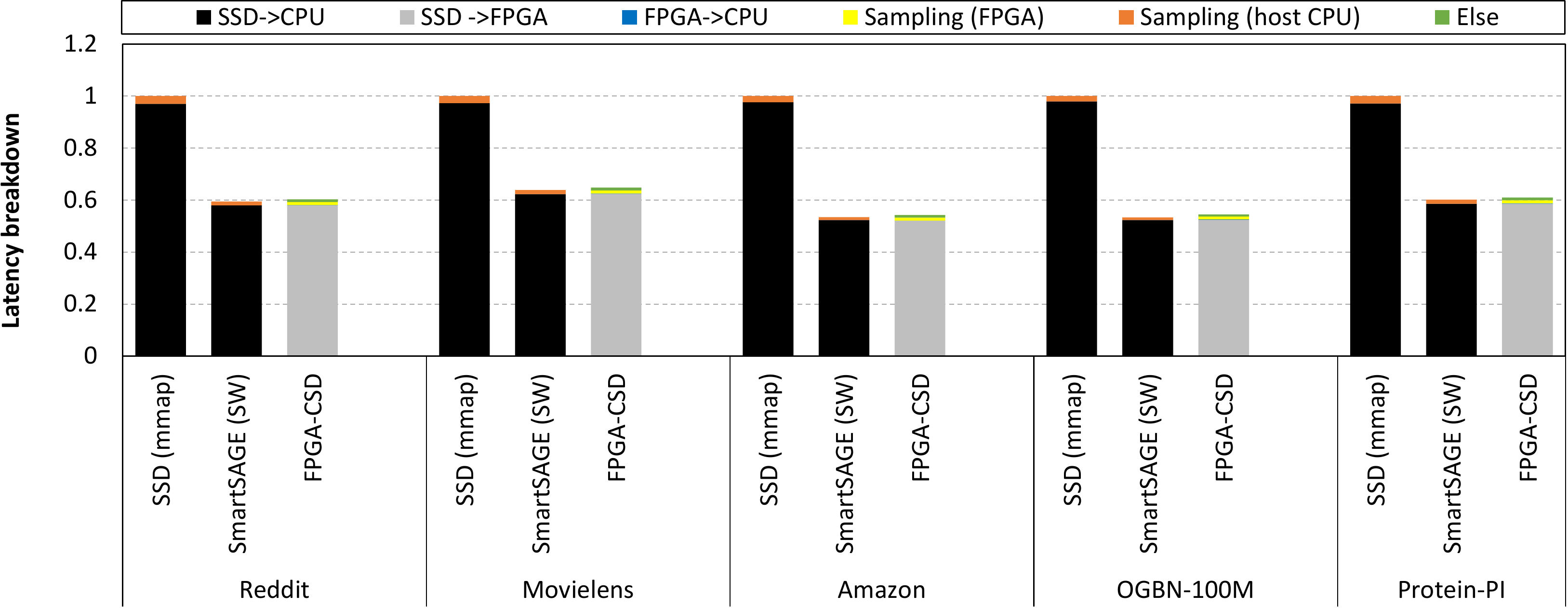}
\caption{
FPGA-based CSD vs. \ssd{mmap} and \sage{SW}.
}
\vspace{-1.2em}
\label{fig:eval_smartssd}
\end{figure}

\subsection{Comparison to FPGA-based CSD Designs}
\label{sect:eval_vs_smartssd}

In \sect{sect:proposed_hw}, we argued that a firmware-based CSD is much more
appropriate for in-storage neighbor sampling compared to an FPGA-based CSD.
\fig{fig:eval_smartssd} quantitatively demonstrates our rationale behind such
design decision. We utilize Samsung-Xilinx's SmartSSD~\cite{smartSSD} to implement a
neighbor sampling operator within SmartSSD's FPGA logic (denoted ``FPGA-CSD'').
Such an FPGA-CSD based design conducts in-storage neighbor sampling by: 1)
copying the necessary chunks of neighbor edge list array from
SSD$\rightarrow$FPGA using P2P transfer (gray), 2) utilizing FPGA's hardwired neighbor node 
gather unit to conduct sampling over FPGA's local DRAM (yellow), and 3) transferring the
sampled subgraph from FPGA$\rightarrow$CPU (blue). As depicted, the performance of FPGA-CSD is bottlenecked on 
the latency to move the neighbor edge list array chunks from SSD$\rightarrow$FPGA, failing to achieve any performance advantage even over our software-only \sage{SW}.

\subsection{Power and Energy Consumption}
\label{sect:eval_energy}

The CPU-GPU based GNN training system consumes several thousands
	of watts of system-wide power (i.e., the CPU, GPU, DRAM, SSD). Due to the following factors,
		 the added power overheads of SmartSAGE
 is expected to be negligible, allowing the significant
 reduction in training time to proportionally
 improve system-level energy-efficiency.
	First, SmartSAGE(HW/SW) is a purely firmware-based CSD that utilizes 
existing embedded cores within the SSD. Consequently,
the added power overheads of SmartSAGE(HW/SW), if any, is amortized by
the significant reduction in end-to-end training time. For the more future-looking
SmartSAGE(oracle) design that integrates additional, \emph{dedicated} ISP-purposed embedded cores (assuming quad core
ARMv8 Cortex-A53 like NGD Newport), the CSD will add around $2-6$ watts of TDP of additional power consumption, which is a reasonable overhead (vs. CPU/GPU/DRAM/SSD's system-level power consumption) given the significant reduction in training time.

\begin{figure}[t!] \centering
\includegraphics[width=0.475\textwidth]{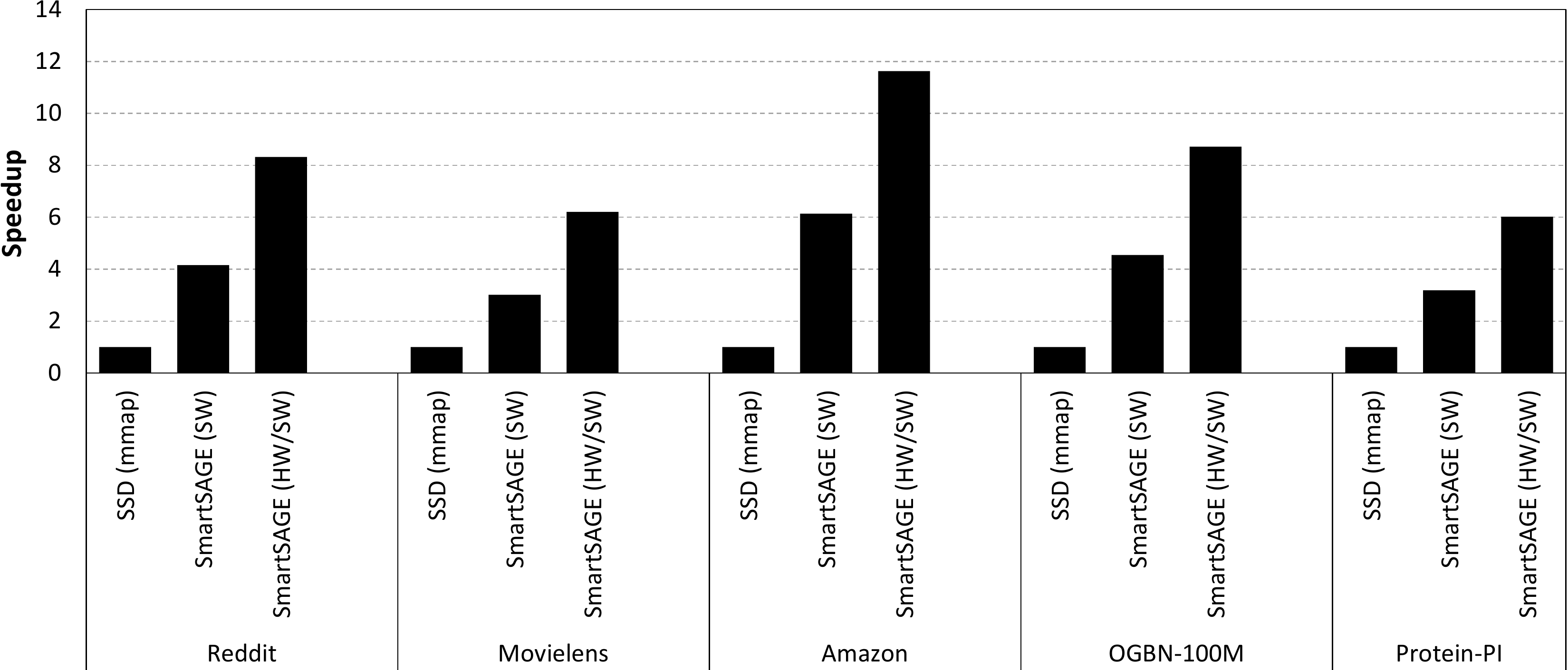}
\caption{
Sensitivity of SmartSAGE's speedup to different sampling algorithm, i.e., GraphSAINT.
}
\label{fig:eval_sensitivity_graphsaint}
\end{figure}
		
\begin{figure}[t!] \centering
\includegraphics[width=0.475\textwidth]{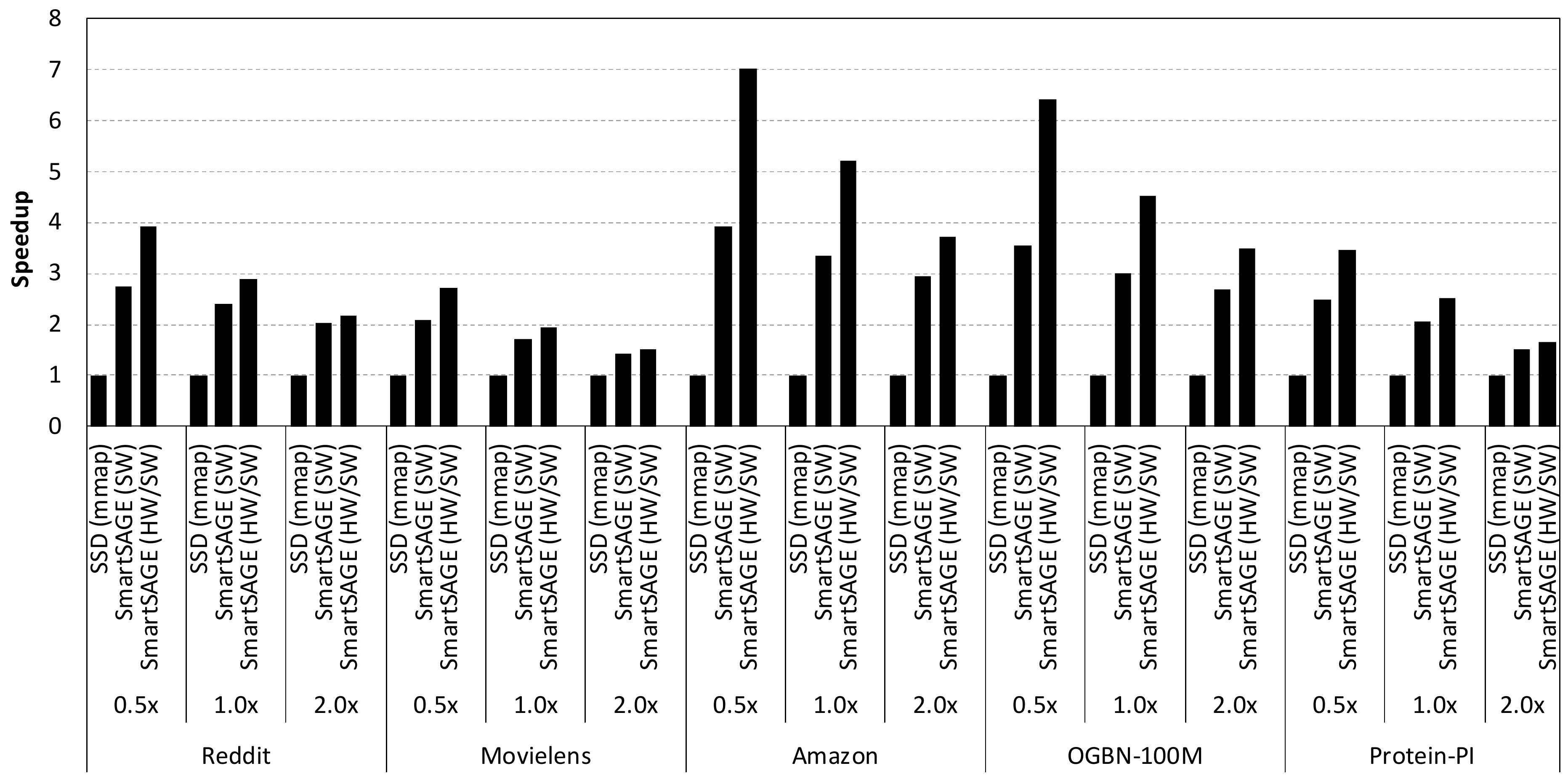}
\caption{
Sensitivity of \proposed's end-to-end speedup to sampling rate.
}
\vspace{-0.5em}
\label{fig:eval_sensitivity}
\end{figure}

\subsection{Sensitivity}
\label{sect:eval_sensitivity}

This subsection evaluates \proposed's robustness to various configuration
parameters of GNN training.

{\bf Alternative sampling algorithm.}
To demonstrate the robustness of our proposal,
	 we implement another	state-of-the-art graph sampling algorithm, GraphSAINT~\cite{graphsaint}.
		Unlike GraphSAGE, GraphSAINT employs a regular random walk based method to sample a neighbor per node across
		multiple target nodes.
		As depicted in
		\fig{fig:eval_sensitivity_graphsaint}, SmartSAGE achieves an average of
		$8.2\times$ end-to-end speedup with GraphSAINT, demonstrating our proposal's
		robustness and flexilibity in accommodating various graph sampling algorithms.
		
{\bf Sampling rate.} \fig{fig:eval_sensitivity} shows \proposed's neighbor
sampling speedup when we sweep the sampling rate by $0.5\times$ and $2\times$
of the default setting (default: $25$ and $10$ neighbor nodes sampling per each
		target node in the first and second GNN layer, respectively).  As depicted,
	 \sage{HW/SW}'s speedup is gradually decreased (increased) as sampling rate
	 gets larger (smaller).  Under \sage{HW/SW}, as we increase the sampling
	 rate, the subgraph being transferred over SSD$\rightarrow$CPU gets larger
	 and eventually approaches the data transfer size of \sage{SW}.
Such phenomenon renders \sage{HW/SW}'s speedup to
	 become smaller than when evaluated under the default sampling rate.

{\bf Training batch size.} We also study \proposed's sensitivity to
larger/smaller mini-batch sizes. Results showed that the chosen mini-batch size
have little effect on \proposed's achieved speedup.  In general,
		we confirmed \proposed's robustness under different design points but omit
		the results due to space constraints.

%% file: tex/discussion.tex
\section{Related Work}
\label{sect:related}

The focus of our work is on accelerating large-scale GNN training with an ISP
architecture.  There is a large body of prior literature exploring
in-storage/near-data processing~\cite{bluedbm, biscuit, active_flash, catalina,
	activesort, morpheus:isca2016, grafboost:isca2018,
	do:query_processing:sigmod13, recssd,
		minsubkim:reducing_tail_latency:apsys2020, willow, summarizer,
			graphssd,recssd,minsubkim:reducing_tail_latency:apsys2020,glist,
			kim2016storage, wang2016ssd, choi2015energy, lee2017extrav,
			xu2020aquoman, hu2019dynamic, kang2013enabling, cho2013active,
			active_disks, idisks, riedel1998active, bae2013intelligent, xsd,
			woods2014ibex} or in-memory
			processing~\cite{tensordimm,recnmp,trim,trim:cal,tensorcasting,samsung_pim,neural_cache,compute_caches,wang:bit_prudent:hpca2019,isaac,prime,alian:mcn:micro2018,chameleon,nda,tetris,neurocube,centaur:hwang}
			architectures for data-intensive workloads as well as ASIC/FPGA/GPU based
			acceleration for graph neural
			networks~\cite{hygcn,engn,awbgcn,gcnax,cambricong,grip,gnnadvisor,graphact,glist,marius,smartsage:cal}.
			There is also prior work exploring heterogeneous memory systems
				for training large-scale ML
                algorithms~\cite{rhu:2016:vdnn,rhu:2018:cdma,mcdla,mcdla:cal,kwon:2019:disagg,capuchin,lms_sysml,swapadvisor,layer_centric:taco,sentinel,superneurons}.
			Due to space limitations, we summarize a subset of these related work
			below.


{\bf ISP designs for data-intensive workloads.}
 BlueDBM~\cite{bluedbm} investigated an ISP architecture targeting big data
 analytics, where a dedicated FPGA fabric is utilized for accelerating nearest
 neighbor search, string search, and graph traversal. Biscuit~\cite{biscuit}
 similarly explores ISP acceleration for big data workloads but the ISP
 substrate utilized is more close to a firmware-based CSD rather than BlueDBM's
 FPGA-based CSD design. Active Flash~\cite{active_flash}, Catalina~\cite{catalina},
 ActiveSort~\cite{activesort}, Morpheus~\cite{morpheus:isca2016},
 GraFBoost~\cite{grafboost:isca2018}, and work by Do et al.~\cite{do:query_processing:sigmod13}
 are also ISP architectures that target data analytics (e.g., pattern matching,
		 $k$-mean clustering, etc), database management, MapReduce frameworks,
 object (de)serialization, graph analytics, and Microsoft SQL, respectively.
 Aside from these prior art
 focusing on domain-specific acceleration of a particular application area,
 there is also a  set of previous work that seeks to enhance the
 programmability of ISP designs. Willow~\cite{willow} explores architectural
 support to enhance the programmability and flexibility of using 
 programmable SSDs for ISP. Summarizer~\cite{summarizer} proposes a set of flexible
 programming APIs for conducting ISP-based data filtering and data
 summarization operations. GraphSSD~\cite{graphssd} proposes a graph
 semantic aware ISP architecture that provides a simple programming interface
 for enhanced programmability.  
The adoption of CSDs for
 domain-specific acceleration is a trait SmartSAGE resembles to these prior
 literature, especially with Biscuit, Summarizer, and GraphSSD. 
Unlike SmartSAGE, however, Biscuit
 employs a dedicated pattern matching IP for ISP,
 which cannot be utilized for GNN's graph sampling.
 Similarly, GraphSSD is specifically optimized for efficient graph data layout mapping
 within the physical pages, focusing on in-storage graph update operators for bandwidth
 amplification, which cannot be readily employed for graph sampling.
 Summarizer bears similarity to SmartSAGE in that it supports in-storage reduction
 operations for generic data-intensive workloads like TPC-H. Because Summarizer does not
 target GNN training, however, it lacks several of SmartSAGE's software-level optimizations
 like direct I/O or mini-batch level I/O command coalescing (\fig{fig:proposed_runtime}).
More crucially, our work targets a new, emerging application domain (i.e.,
		 machine learning based graph neural network training) and uncovers a new
 system-level bottleneck driven by different
 intuitions, rendering our key contributions unique.

{\bf Accelerating GNN inference.} 
	There is also a rich set of previous work exploring hardware/software acceleration techniques for
machine learning inference~\cite{hygcn,engn,awbgcn,gcnax,cambricong,grip,gnnadvisor,recssd,minsubkim:reducing_tail_latency:apsys2020}.
While not specifically targeting GNNs, RecSSD~\cite{recssd} and work by Kim et
al.~\cite{minsubkim:reducing_tail_latency:apsys2020} are recent ISP designs utilizing 
firmware-based CSDs to overcome the memory capacity bottlenecks of the
inference stage of DNN-based recommendation models.
HyGCN~\cite{hygcn} is one of the first ASIC based GCN accelerators providing
substantial energy-efficiency improvement for the backend GNN layers of graph
learning. GLIST~\cite{glist} is an FPGA-based CSD for GNN inference and GNNAdvisor~\cite{gnnadvisor}
proposes an efficient software runtime for GPU-based GNN execution.
Unlike these prior
work focusing on accelerating the inference of GNNs, \proposed focuses on 
uncovering
the system-level bottlenecks of training, more specifically the frontend data
preparation stage.  Overall, the key contribution of \proposed is orthogonal to
these prior studies.

{\bf Accelerating GNN training.} GraphACT~\cite{graphact} proposes a CPU-FPGA
based acceleration platform for GNN training. Marius~\cite{marius} is a software
framework that aims to provide high-performance large-scale graph learning within a single machine.
Unlike \proposed however, Marius targets a different graph learning algorithm so the model
parameters trained with Marius are the graph embedding vectors, not
the GNN layers (as is the case with \proposed). More importantly,
Marius proposes a software-level pipelining solution for fast training, unlike the ISP based \proposed.
Similar to SmartSAGE, prior work on AliGraph/PaGraph/DistDGL~\cite{alibaba,pagraph,distdgl}
employ off-the-shelf
	graph partitioning algorithms for large-scale GNN training, partitioning the graph
	sampling and learning process
	across multiple workers for parallel GNN training. As pointed out by Su et al.~\cite{gnnsys_21}, distributed GNN
	training can suffer from load imbalance issues (e.g., number of $k$-hop sampled nodes
	can	differ across graph partitions),
	intermittent model parameter synchronization overheads, and etc.
SmartSAGE explores a different research space for GNN training systems, enabling
large-scale GNN training within a single machine using NVMe SSD while still achieving DRAM-level
performance.
Overall, the
contribution of \proposed is orthogonal to these studies.

%% file: tex/conclusion.tex
\section{Conclusion}
\label{sect:conclusion}

In this work, we investigate the viability of utilizing NVMe SSDs to overcome the memory capacity limitations of current, in-memory
processing GNN training systems. 
We propose an in-storage processing based GNN training system called \proposed which
synergistically combines the ISP capabilities of emerging CSDs
with a latency-optimized software runtime and host driver stack. By intelligently
offloading the data intensive frontend data preparation stage of GNN training,
\proposed significantly resolves the bottlenecks of the baseline SSD-centric training
system, achieving significant performance improvements.